\PassOptionsToPackage{hyphens}{url}

\documentclass[10pt,letterpaper]{article}
\usepackage[top=0.85in,left=2.75in,footskip=0.75in]{geometry}

\usepackage{amsmath,amssymb}

\usepackage{changepage}

\usepackage[utf8x]{inputenc}

\usepackage{textcomp,marvosym}

\usepackage{cite}

\usepackage{hyperref}
\usepackage{nameref}

\usepackage{microtype}
\DisableLigatures[f]{encoding = *, family = * }

\usepackage[table,dvipsnames]{xcolor}

\usepackage{array}

\usepackage{multirow}
\usepackage{tabularx}
\usepackage{array, makecell}

\newcolumntype{+}{!{\vrule width 2pt}}

\newlength\savedwidth

\usepackage{setspace} 
\raggedright
\usepackage[aboveskip=1pt,labelfont=bf,labelsep=period,justification=raggedright,singlelinecheck=off]{caption}

\makeatletter
\renewcommand{\@biblabel}[1]{\quad#1.}
\makeatother

\usepackage{graphicx}
\usepackage{lastpage,fancyhdr}
\usepackage{epstopdf}
\fancyheadoffset[L]{2.25in}
\fancyfootoffset[L]{2.25in}
\DeclareGraphicsExtensions{.pdf,.eps,.png,.jpg,.tif,.tiff,.ps}
\graphicspath{{./figures/}{./figures/PDF-versions/}}
\usepackage[nameinlink,capitalize]{cleveref}
\usepackage{booktabs}
\usepackage[colorinlistoftodos,textsize=tiny]{todonotes}
\reversemarginpar
\usepackage{soul}
\usepackage[hyphens]{url}

\definecolor{navy}{rgb}{0.1, 0.1, 0.8}
\definecolor{gray}{rgb}{0.4, 0.4, 0.4}
\definecolor{olive}{rgb}{0.1, 0.5, 0.1}
\definecolor{ruby}{rgb}{0.8, 0.1, 0.3}

\newcommand{\eat}[1]{}

\begin{document}
\vspace*{0.2in}

% Title must be 250 characters or less.
\begin{flushleft} 
{\Large
\textbf\newline{Evolution of diversity and dominance of companies in online activity % Please use "sentence case" for title and headings (capitalize only the first word in a title (or heading), the first word in a subtitle (or subheading), and any proper nouns).
}
%\dsf{Shorter title ok?}

% Insert author names, affiliations and corresponding author email (do not include titles, positions, or degrees).
Paul X. McCarthy\textsuperscript{1,*},
Xian Gong\textsuperscript{1},
Sina Eghbal\textsuperscript{2}, 
Daniel S. Falster\textsuperscript{3}, and
Marian-Andrei Rizoiu\textsuperscript{4}.
\\
\bigskip
\textbf{1} School of Computer Science and Engineering, University of New South Wales, Sydney NSW 2052, Australia
\\
\textbf{2} College of Engineering and Computer Science, The Australian National University, Canberra, Australia
\\
\textbf{3} Evolution \& Ecology Research Centre, School of Biological, Earth and Environmental Sciences, University of New South Wales, Sydney NSW 2052, Australia
\\
\textbf{4} UTS Data Science Institute, University of Technology Sydney, Sydney NSW 2007, Australia
\\
\bigskip

% Insert additional author notes using the symbols described below. Insert symbol callouts after author names as necessary.
% 
% Remove or comment out the author notes below if they aren't used.
%
% Primary Equal Contribution Note
%\Yinyang These authors contributed equally to this work.

% Additional Equal Contribution Note
% Also use this double-dagger symbol for special authorship notes, such as senior authorship.
%\ddag These authors also contributed equally to this work.

% Current address notes
%\textcurrency Current Address: Dept/Program/Center, Institution Name, City, State, Country % change symbol to "\textcurrency a" if more than one current address note
% \textcurrency b Insert second current address 
% \textcurrency c Insert third current address

% Deceased author note
%\dag Deceased

% Group/Consortium Author Note
%\textpilcrow Membership list can be found in the Acknowledgments section.

% Use the asterisk to denote corresponding authorship and provide email address in note below.
* paul.mccarthy@unsw.edu.au
}
\end{flushleft}

%!TEX root = ../main.tex
%
% Please keep the abstract below 300 words
\section*{Abstract}
Ever since the web began, the number of websites has been growing exponentially.
These websites cover an ever-increasing range of online services that fill a variety of social and economic functions across a growing range of industries. %~\cite{Albert1999,Huberman1999,Gandhi2016}
Yet the networked nature of the web, combined with the economics of preferential attachment, increasing returns and global trade, suggest that over the long run a small number of competitive giants are likely to dominate each functional market segment, such as search, retail and social media. %~\cite{Salganik2006,McCarthy2015,Scheffer2017,moore2018digital}.
Here we perform a large scale longitudinal study to quantify the distribution of attention given in the online environment to competing organisations.
In two large online social media datasets, containing more than 10 billion posts and spanning more than a decade, we tally the volume of external links posted towards the organisations' main domain name as a proxy for the online attention they receive. 
We also use the Common Crawl dataset -- which contains the linkage patterns between more than a billion different websites -- to study the patterns of link concentration over the past three years across the entire web.
%We also analyse broader patterns of link concentration over the past three years across the entire web, using open data about linkage patterns relating to over 1 billion different websites.
%\mar{Brought the case study after the main findings.}
Lastly, 
%we take a preliminary look at potential future research directions in 
we showcase the linking between economic, financial and market data by exploring the relationships between online attention on social media and the growth in enterprise value in the electric carmaker Tesla.%

Our analysis shows that despite the fact that we observe consistent growth in all the macro indicators -- the total amount of online attention, in the number of organisations with an online presence, and in the functions they perform -- we also observe that a smaller number of organisations account for an ever-increasing proportion of total user attention, usually with one large player dominating each function. These results highlight how evolution of the online economy involves innovation, diversity, and then competitive dominance. 
%

% \linenumbers
%!TEX root = ../main.tex
%
% Use "Eq" instead of "Equation" for equation citations.
\section*{Introduction}

%% first, talk about the emergence of giants
\paragraph{A brave, new online world.}
%\paragraph{The rise of online giants.}
Although now over two decades old, the web remains a relatively young platform for economic and social interactions, with new functionalities and possibilities continuously arising. 
Early views of the web saw the online economy as an open platform that would encourage a flourishing of diverse interests~\cite{friedman2005world,levine2009cluetrain}. 
However, as activity on the web has grown, attention has focused more intently on how different organisations compete for the limited attention and resources of web users~\cite{da2011search,xiang2010role}.
Several theories suggest that particular features of the web---such as positive network effects, modest switching costs and dissolving of geo-political boundaries for competition---naturally favour the emergence of online giants~\cite{McCarthy2015,Scheffer2017,shapiro1998information}. 
While the very existence of large internet platform companies provides anecdotal support for the competitive dominance in the digital economy, the lack of consistent, long-term data makes it difficult to see how a small number of companies have emerged as globally dominant in their respective functional domains (such as Amazon in retail, Google in search, and Facebook in social media).

The online and offline economies are intimately linked, and the economic forces that have helped fuel the growth of online giants have broader effects across the whole offline economy. 
For example, the fact that industries become more digitised and more connected makes them more profitable, but it also stifles competition and shields the leading firms from new rivals.
This may be partially explained by these companies monopolising the attention of online users---who map back to real offline consumers---which in turn has been shown to result in an increasing dominance of a small number of firms, who are more profitable and face less new entrants~\cite{wang2015role}.
Online attention has been previously measured through metrics such as counts of online searches~\cite{Danescu-Niculescu-Mizil2010}, video views~\cite{Rizoiu2017,Wu2019}, retweets~\cite{Mishra2016,Zhang2019} or webpage links~\cite{wang2015role}, and has been shown to predict patterns of offline behaviour~\cite{jun2014study} accurately.
For example, the volume of specific web search terms has been shown to correlate to the real world social and economic phenomena such as unemployment~\cite{d2017predictive}, housing price trends~\cite{wu2015future} and the relative market shares of sales of companies in emerging markets such as electric car brands~\cite{jun2014study}. 
Search traffic on specific terms~\cite{ginsberg2009detecting}, visit counts~\cite{mciver2014wikipedia} and edits made~\cite{Rizoiu2016} to specific Wikipedia articles and terms used in Twitter posts have also been shown to correlate to population dynamics (for example, the spread of influenza~\cite{aramaki2011twitter}). % PXM - suggest we use population or social dynamics instead of human dynamics % 
As always, caution is needed in interpreting these results, as it is impossible to demonstrate causative links from observational data of large complex systems. Yet few would deny that a company's online presence could be causally linked to its long-term success.
% Growth in social media attention may be a consequence of giants' growth and not a principal cause %% MAR: not sure about this one, as we show the Granger-causality for Tesla EV

%% second, emergence of niches or functions
%\textbf{The emergence of functions.}
New commercial and social uses for the web --- dubbed here as \emph{functions} --- are continually being explored and developed.
Often new technologies enable such new functions to emerge. 
For example, the advent of secure data communication on the web set the stage for e-commerce and for companies like Amazon to emerge; 
the roll out of broadband-enabled video enabled applications such as Youtube and Netflix to grow; further, 
the democratisation of mobile terminals (i.e. smartphones) gave rise to location-based applications such as Uber and Airbnb. 
Peer-to-peer accommodation sharing, ride sharing and ephemeral social media are three of the recent innovative functions. 
%% third, competition inside functions
As new functions emerge online and a field of new competitors assemble, the relative growth in companies' revenue in a function can indicate which company will come to dominate (and which can be under some circumstances modelled and predicted~\cite{kutcher2014grow,lera2020prediction}).
Revenue in turn is a function of growth and highly dependent on the retention of customers~\cite{gupta2004valuing} and, particularly, their attention~\cite{bozovic2017unicorns}. 
%\mar{Closed the argument towards why we look at attention.}
As a result, the early distribution of online attention within a function can be seen as a leading indicator of the company that will dominate the function.

Between 2006 and 2017, the web has grown significantly in terms of its number of users, average usage time by the users and the number of organisations active online.
For example, 
%\mar{I know non-CS audience frowns upon usage of bullet points. Linearizing the text.}
%, how much each person uses the web and how many organisations are active on the web, namely:
%\begin{itemize}
%  \item 
the number of internet users worldwide has grown from 1.2 Billion 2006 to 3.3 Billion in 2016~\cite{ITU2020}.
%  \item 
Similarly, the amount of time spent online grew steadily over the years, from 2.7 hours per day in 2008 to 5.7 hours to per day 2016 in the US~\cite{BondInternet2019};
%  \item 
Finally, the number of organisations, brands and services represented online has grown dramatically. 
  The total number of domain name registrations grew from 79 million in 2006~\cite{Internet2006} to 329 million in 2016~\cite{Verisign2016}.
%\end{itemize}
%
Network theory shows that the web grows following the law of increasing returns~\cite{arthur1989competing} where new links are added where others already exist (also known as power-law or preferential attachment, also shown in our data, see \nameref{sec:materials-methods}).
This phenomenon leads to a cumulative advantage for a small number of organisations (and their online services) that enjoy most of the attention, while the others attract very little~\cite{barabasi1999emergence}. 
Network topography has also been shown to govern user attention and activity, which also follow a power-law distribution~\cite{watts2002simple}, as does the number of users across websites relative to their rankings by total visits, and the total attention given to each of these websites~\cite{adamic2002zipf}. 

%\paragraph{The rise (and fall) of online businesses.}
%In this work,
We leverage the above-mentioned
%Extending these  from 
network theory insights, combined with a novel data source (see \nameref{sec:materials-methods}), to study the rise and fall of attention given to businesses online. To quantify the dynamics of activity online, we present a longitudinal study of user attention on two popular social media sites --- Reddit and Twitter. 
Our analysis spans over 10 years of Reddit history and more than 6 years of Twitter history. 
%
%As a proxy for online attention, 
Both Reddit and Twitter allow users to share ideas and links on topics of interest, with 310 million and 328 million global users respectively~\cite{Waters2017Reddit}. 
As such, these public platforms provide a chronicle of internet users' attention and interests over time.
Note that we do not follow individual users, we measure the aggregated attention patterns \emph{towards} companies.
We use the volumes of outbound links posted to a companies' website as a proxy for user attention towards the organisation and its products.
%\verify{We use the outbound linkage patterns on these two platforms as proxies for how users allocate their attention to different online organisations. }

Employing weblinks has several advantages. First, they are structured around domain names. 
Usually, each domain corresponds to a single company---in this work, we use \emph{domain} to interchangeably denote a web domain or a company with an online presence.
Second, weblinks are easily identifiable from the surrounding natural language text and easily tallied. 
Third, links are central to the web's architecture, and link counts are significant indicators of the website quality and authority.
They are also at the heart of the ranking in search indexes, including Google's PageRank algorithm~\cite{page1999pagerank}.
%\mar{@Paul: We would need to discuss here also the addition of Common Crawl, and how it confirms the findings from the social media data.}
In fact, our present work shows that the PageRank of domains and the online attention quantified from social data are distributed very similarly (see \nameref{sec:results} and \cref{fig:fig2}). 
In an increasingly global digital economy, online attention is a new form of currency. 
One key measure is online links, which are shown to drive attention which in turn drives online dominance which, we postulate, is linked to market dominance. 

There are several reasons why unconventional data, such as social media links, are a beneficial and useful source of information for researchers interested in business dynamics.
In addition to being publicly available and timely (and massive in scale) these data have at least three advantages over traditional financial market data. 
First, for new and emerging categories, traditional financial measures are often unavailable or not meaningful. 
Second, sometimes the direct competitors are not known.
For examples, the competitors are often private companies part of broader business conglomerates with multiple revenue streams; therefore, individual functional business unit financial data is not available.
Third, in some cases, especially in emerging and new industries such as electric vehicles, trends in social media data may precede, anticipate or even predict trends in businesses, financial markets and the broader economy.  

\subsection*{Competitive diversity and its role in economics.}  
In Economics, a variety of independent firms competing with each other in each market and segment of the industry is vital for the economy's health.
Diversity is at the heart of all effective competition and, in turn, has been shown to lead to higher rates of productivity growth~\cite{pop00001}.
The number, structure and variety of competing firms in a sector in Economics can be seen as analogous to the diversity of species in Ecology, in a niche. 
%In this study, we measure the diversity of online firms using a measure shared by Ecology and Economics.
%%Our measure of diversity of firms online in this study is one that is shared with Ecology. 
%In economics it is known as Herfindahl-Hirschman Index (or HHI) wheras in ecology it is known as the Simpson Index.
%
If there are too few competitors or a small number of players become too dominant within any economical sector, there emerges the potential for artificially high prices (monopoly rent) and constraints to supply.
Even more importantly, in the long-term, this gives rise to constraints on innovation. 
Nobel-winning economist Ken Arrow~\cite{arrow1962economic} postulated that, once markets are dominated by an established firm or group of firms, this establishment has less incentive to innovate.
This is because they would have an added cost to innovation that an innovating competitor would not --- the opportunity to continue to earn monopolistic profits without innovating.

We explore the large-scale and long-term trends in industry structure and economic variety through the lens of online attention in several key dimensions, including:
\begin{itemize}
  \item scale---how many different organisations are active online;
  \item originality---how many distinctive sources of information and services are there online;
  \item diversity---how attention is divided and distributed between firms online.
\end{itemize}

We also examine online innovation through the birth of new online business functions, such as ride-sharing, online video, and ephemeral peer-to-peer messaging.
We propose that competition among organisations online follows a three-phase dynamic~\cite{McCarthy2015}:
\begin{enumerate}
   \item Infancy---after the emergence of a new function, we observe a great burst of diverse businesses that appear and start to serve the function;
   \item Development---this phase begins once the number of competitors within the function peaks and begins to dwindle, as competitors start to lose market share to one another;
   \item Maturity---during which we see a reduced diversity in the function, as the majority of users converge around a single dominant organisation.
\end{enumerate}

%\paragraph{}
%\paragraph{Measuring online attention towards companies.}
%In this work, 
%\paragraph{Patterns of activity online represents economy-wide trends.}  

%!TEX root = ../main.tex
%
% Use "Eq" instead of "Equation" for equation citations.
\section*{Materials and Methods}
\label{sec:materials-methods}

%In this section, we describe the technical work and the employed datasets.
%We analyse and interpret results in Sec.~"\nameref{sec:results}".
%
%\subsection*{A window into a decade of web activity}
%
%\paragraph{Methodology.}

In this work, we longitudinally integrate and analyse three types of data sources: 
1) proxies for online attention towards companies, 
2) the list of competitors for online companies and 
3) the economic performance of companies.
We measure \textbf{the attention towards companies online} by examining links posted over a decade from two major online platforms: Reddit (\url{www.reddit.com}) and Twitter (\url{www.twitter.com}).
We use social media data as a proxy for online human attention towards companies.
We also use Common Crawl (\url{https://commoncrawl.org/}), which records the PageRank~\cite{page1999pagerank} over time, and which can be seen as a proxy for the attention received from internet websites.
We combine and index these datasets with two other data sources which record \textbf{competitor data}: Crunchbase (\url{https://www.crunchbase.com/}) and Rivalfox (which closed in 2017, but for which the competitor maps are still available via the Internet Archive---e.g., for Airbnb competitors from Rivalfox see \url{https://web.archive.org/web/20150327025122/https://rivalfox.com/airbnb-competitors}).
Finally, we measure the companies' \textbf{economic performance} by building the \emph{enterprise value} using historical data from Financial Times (\url{https://www.ft.com/}) and Yahoo Finance (\url{https://finance.yahoo.com}).

\subsection*{Data collection}
Here we discuss the collection of online attention data (social media and webpage linking) and economic performance data.
All the constructed longitudinal datasets and all the code required to reproduce the research and the figures in this paper are publicly available online at \url{https://github.com/behavioral-ds/online-diversity}.
%We examine large-scale, cross-sectional, longitudinal patterns in , combined and indexed with three other online
%platforms Crunchbase (\url{https://www.crunchbase.com/}), Rivalfox (which closed in 2017, but for which the competitor maps are still available via the Internet Archive\footnote{Airbnb competitors from Rivalfox: \url{https://web.archive.org/web/20150327025122/https://rivalfox.com/airbnb-competitors}}) and Common Crawl (\url{https://commoncrawl.org/}).

\textbf{Social media data.}
From Reddit, we used the publicly available dumps of Reddit comments (see \url{http://files.pushshift.io/reddit/comments/}), which are claimed to capture all Reddit activity from December 2005 until December 2019.
Each dump file contains all the comments posted on Reddit during one month, in JSON format.
The available fields include the posting date, the author and the text of the comment.
We compiled and analysed more than 10 years worth of data, comprising more than 6 billion Reddit user comments.
For Twitter, we used data from a long-running crawler leveraging the Twitter Sampled stream (retired as of Oct 2020, see \url{https://developer.twitter.com/en/docs/labs/sampled-stream/overview}), which returns in real-time a sample of 1\% of all public tweets.
The crawler ran continuously from September 2011 until the end of 2019, and the datasets contains several gap periods (seen in \cref{{fig:fig1}} and the Supporting Information), when network errors took the crawler down.
%Twitter ToS does not allow us to share this data in its raw format.
We compiled and analysed more than 11.8 billion user posts (known as tweets), published from September 2011 to September 2019 (see \cref{tab:data-dimension} for datasets stats). 
%We use social media data as a proxy for online human attention towards companies.

%\verify{
%Original reddit data was from 3rd party source jason baumgartner of http://pushshift.io, which claims to contain a dump of all Reddit activity. We use the “comments” file, which contains the text of all comments posted on Reddit. The data dump contains data from Dec 2015 until the end of 2019.
%
%For Twitter, we used a long-running data crawler using the 1\% Streaming API\footnote{Twitter Sampled stream (retired as of Oct 2020): \url{https://developer.twitter.com/en/docs/labs/sampled-stream/overview}} without any query. The Twitter API help pages state that this returns 1\% of all tweets posted on Twitter. The crawler has data from Sept 2011 until the end of 2019. Note that this data source contains gap periods, where errors took the crawler down. Twitter ToS does not allow us to share this data in its raw format.

%Both Reddit posts and tweets were filtered to only those that contain an URL.}

%\TODO{MAR}{Add more info about how Twitter was crawled!}
%
%\verify {====================================}

\textbf{Common Crawl web linking data.}
We compile the linkage record over time of online websites towards domains using Common Crawl (\url{https://commoncrawl.org/}).
Common Crawl is the largest open index of the web and which has been shown to represent over 80 per cent of the world's most popular public web sites when compared with data from Amazon (Alexa) and other sources~\cite{CommonCorpusCoverage}.
The raw data record a large array of linkage features for each quarter between May 2017 and May 2020.
The 69 million most linked domains, Common Crawl computes the PageRank for each domain, which we use as a proxy for the attention received from internet websites:
the higher the PageRank, the more central a domain is and therefore the more attention (weblinks) it receives.

%
%
%\verify{Additionally, we analyse linkage trends across the entire web over the last three years (May 2017 -- May 2020) using data in the Common Crawl (\url{https://commoncrawl.org/}) the largest open index of the web and which has been shown to represent over 80 per cent of the world's most popular public web sites when compared with data from Amazon (Alexa) and other sources~\cite{CommonCorpusCoverage}. 
%
%From Common Crawl data we examine the market share trends in web linkages over time using Quarterly datasets containing pagerank values and positions/rank (known web centrality ranks of web pages) of over 69 million domains. 
%As the PageRank values of all domains in each period add to one, summing the PageRanks of the top 1k, 10k and 1 million domains in each period forms an effective measure for relative market share of total inbound links to each of these cohorts over time. Even though the membership or which domains are represented in these cohorts will change, by looking at the top n domains in each period reveals the broader pattern of link concentration.}
%
%We use web linking data as a proxy for the attention receive from internet websites.

\textbf{Economic performance data.}
For one case study, we aim to examine the relationship between the enterprise value of electric vehicle manufacturer Tesla and the online attention that it receives.
We collected time-series data relating to publicly listed electric car maker Tesla, Inc (NASDAQ:TSLA) between December 2015 and September 2019.
The enterprise value for Tesla was sourced by using the current market capitalisation for Tesla from Financial Times (\url{https://www.ft.com/}) in USD and we used the adjusted historical time-series share price data (Yahoo Finance, \url{https://finance.yahoo.com}) to determine historic enterprise value of Tesla, adjusted for both dividends and splits.

%\paragraph*{S1 Table}
\begin{table}[htbp]
	\caption[Summary of the dimensionality of our datasets.]{
		\textbf{Summary of the dimensionality of our datasets:}
		the number of posts, links contained in posts, and unique domains linked in each of the two datasets.
		For reference only, we also show the size of the dataset in occupied disk space (both datasets are compressed with the BZIP2 utility).
		The combined number of domains (indicated by *) is the number of unique domains in our datasets.
	}
	\centering
	\begin{tabular}{lrrrr}
		\toprule
		Dataset & \multicolumn{1}{l}{\# posts} & \multicolumn{1}{l}{\# links} & \multicolumn{1}{l}{\# domains} & \multicolumn{1}{l}{Dataset size (TB)} \\ \midrule
		Reddit & 6,095,691,657 & 353,915,452 & 4,377,343 & 0.34 \\ 
		Twitter & 11,813,340,769 & 1,553,959,383 & 10,152,547 & 5.25 \\ 
		Common Crawl & - & 19,989,755,161 & 90,983,688 & 0.03 \\ 
		Combined & 17,909,032,426 & 21,897,629,996 & 95,709,245* & 5.62 \\ 				\bottomrule
	\end{tabular}
	\label{tab:data-dimension} 
\end{table}

%In particular, we count the number of distinct domain names contained within posts on each social media platform over each month.
%%we counted the number of posts, the number of links contained within posts and the number of active domains these links referred to. 
%Next, we count the number of outbound links contained within them
%%, for example to domains such as \url{youtube.com}, \url{vimeo.com}, or \url{dailymotion.com}. 
%and we consider a domain as being active if it is linked to at least once within an observation period. 

%\subsection*{Data availability}
%We have created a repository \url{www.github.com/behavioral-ds/online-diversity}
%with all key data for the research including:
%\begin{itemize}
% \item{Reddit longitudinal data}
% \item{Twitter longitudinal data}
% \item{ Common Crawl longitudinal data}
% \end{itemize}
%As well as all code for recreation of figures.

\subsection*{Measuring the spread of online attention}
Here, we further detail the measurements we perform to quantify the dynamics of online attention allocation.
First, we look at the distribution of online links in social media and model their best fit.
Next, we investigate its concentration over time using several indicators.
Finally, we propose a simple measure of \emph{link originality} to measure online diversity, and we examine its evolution over time.

%For both Reddit and Twitter, we tabulated the outbound links over time within the social media posts.  
For both Reddit and Twitter, we tally the number of posts and outbound links within user posts during observation periods of one month.
Outbound links point to websites and online services outside the host platform.
Next, we classify each link based on its major domain --- for example, a link such as \url{https://www.youtube.com/watch?v=dLRjiiAawGg} refers to the domain \url{www.youtube.com}, which belongs to the online video hosting platform Youtube.
We consider a domain as being active if it is linked to at least once within an observation period.
%% MAR: we moved discussion of results in Results.
%, and we show the number of active domains on each platform over time.

\paragraph{Measures of attention diversity.} 
To understand and quantify patterns and trends in the diversity of online businesses, we use a measure that is common to both Ecology and Economics --- known as the Simpson Index~\cite{simpson_measurement_1949} in Ecology, and the Herfindahl-Hirschman Index (or HHI)~\cite{hirschman1964paternity} in Economics.
We use HHI to measure the diversity of online domains based on the number of links that point to them.
Formally
%
%An alternative measure of attention concentration is the Herfindahl-Hirschman Index (HHI), widely used in economics and ecology:
\begin{equation*}
	HHI = \sum_{i=1}^N S_i^2
\end{equation*}
where $S_i$ is the percentage of all the links within a platform relating to domain $i$, and $N$ is the number of distinct domains at each period in time. 
%The HHI is used as an indicator of the degree of market dominance of the companies within a market or segment, ranging from 0 (evenly distributed market share) to 1 (complete monopoly).
HHI values range from zero to one. The higher the HHI value the less evenly distributed the links are. HHI values that are very low and close to 0 represent when links are evenly distributed between domains (perfect competition is achieved when all firms have equal shares).
Conversely, high HHI values represent a very uneven distribution of links (for complete monopoly, the HHI is 1 when one firm attracts all the links).
Due to the size of our datasets, HHI was computed on a cluster computing environment using the R programming language~\cite{RCore}, using the package \texttt{DescTools}.
%% MAR: not needed anymore, as Fig 1C shows the HHI for the whole Twitter.
%Note that for Twitter, to filter for automated retweeting and delayed tweeting services and robotic tweets that account for a material portion of tweets, we have constrained HHI to measure links to the Top 1000 most popular sites on the Web (measured using Quantcast Global Ranking, for more details see the Supporting Information), and we excluded self-citations to \url{twitter.com}.

We also propose a new measure to quantify the bias in the distribution of attention -- dubbed \emph{link originality} -- defined as the average number of domains per link. 
Link originality takes values between $\approx 0$ (absolute monopoly, all links stem from the same domain) and $1$ (complete diversity, each link has a distinct and original domain).

The logic behind link originality is as follows.
A known measure for the skewness of a distribution is the difference between the mean and the median value. 
Even though some domains feature millions of links per month, half of all domains in our datasets are linked at most once each month -- i.e., the median number of links per domains is one (1) for each analysed time frame. 
Therefore, we can measure the increasing skewness of the attention distribution by tracing the average number of links per domain comparatively to the fixed median of one. 
Link originality is defined as the ratio of number of domains to the number of links, and it is the inverse of the mean attention received by domains. 
Link originality is a simple, intuitive measure for online diversity: when originality decreases, mean domain attention increases, the difference between mean and median attention increases, indicating that the distribution is increasingly skewed.

\paragraph{Measures of attention concentration using PageRank and CommonCrawl.}
We measure the spread of the ``attention of webpages'' using the CommonCrawl dataset, which records the PageRank for each domain.
PageRank quantifies the inbound links to a page to determine how important a website is, assuming that more important websites are likely to receive more links from other websites.
The PageRank for all domains in each period adds to one -- an intrinsic property of the PageRank.
Summing the PageRanks of the \emph{top $n$} domains in each period forms an effective measure for the relative market share of total inbound links to each these `important domains' over time. 
Even though the composition of \emph{top n} will change over time, their total PageRank in each period can reveal the broader trends of link concentration.

\subsection*{Linking social media attention and enterprise value}
To explore, whether there is a link between attention on online social media platforms towards particular companies and their enterprise value, we use the time-series financial market data for the electric vehicle manufacturer Tesla, and the count of link counts posted on Reddit and Twitter towards its domain. 
We chose the Tesla case study given that previous research has shown that a growth in links in social media is predictive for the growth in sales and market share of electric car brands~\cite{jun2014study}.
%% MAR: we never test this
%especially for new categories of products, industries and functions -- 

We use statistical analysis to explore whether a growth in links on social media is predictive of a growth in enterprise value for Tesla. 
We use three time-series data relating to publicly listed electric car maker Tesla, Inc (NASDAQ:TSLA) between December 2015 and September 2019.
The first series is the Enterprise Value (EV) of Tesla.
The second and third series are the counts of outbound links to \url{tesla.com} on Reddit and Twitter, respectively.
We perform the analysis in two steps.
\textbf{In the first step,} we perform stationarity tests to see whether the mean, variance, and autocorrelation for each time series are stable over time.
We use the augmented Dickey-Fuller test (ADF), and we obtain that only the Twitter series appears stationary (p-value $< 0.05$).
We transform all series by \emph{differencing} -- i.e., compute the differences between consecutive observations -- which renders them stationary.
Next, we perform co-integration tests to estimate the long-term equilibrium of two series in order to rule out the possibility of spurious correlation.
We obtain that none of the two pairs (Reddit and EV, Twitter and EV) are co-integrated (see the \nameref{S3_Appendix} in the supplementary information for more details).

\textbf{In the second step}, we examine whether links in these social media data sets (Reddit and Twitter) can be used to forecast the changes of enterprise value using Granger-causality. 
We vary the lag parameter in the Granger-causality test between 1 and 12 periods (corresponding to 1 to 12 months), and we obtain that for a number of lags larger than two (for Reddit) and four (for Twitter), the granger-causality test is significant (p-value $< 0.05$).
Finally,  we use vector auto-regression (VAR) to determine the optimum number of lags by selecting the value which minimises the Akaike information criterion (AIC).
We obtain that the optimal lag is two for Reddit and four for Twitter.
Further details of each of these steps and their results are included in the Supporting Information.

%\verify{
%Once gathered, we performed a number of standard statistical tests on these data to understand if and how they are temporally and causally related. 
%
%First, we performed both stationarity and co-integration tests for each of these data to see whether the mean, variance and auto-correlation are stable over time and to estimate the long-term equilibrium of two series in order to rule out the possibility of spurious correlation. 

%Then, we examined whether links in these social media data sets (Reddit \& Twitter) can be used to forecast the changes of enterprise value using Granger causality. 
%
%Then, lastly we used vector auto-regression (VAR) to explore whether posts on news-centric Twitter may indeed be a leading indicator for trends in growth in investors re-evaluation of Tesla's value by looking at tests for various lags between counts of links to Tesla.com Twitter and subsequent changes in its Enterprise Value.  

%}

\subsection*{Categorising new functions and innovation in the online economy}
\label{subsec:attention-economy}

%\paragraph{The emergence of new online functions.}
The widespread adoption of key digital platform technologies such as security, mobility and broadband themselves enable waves of new business opportunities to emerge. Platform technologies create the conditions to offer services that could not be offered previously for example when security was added to the web it moved from being an information medium to becoming a transaction medium and, in the process, enabled a plethora of new commercial services such as online retail, online payments and online banking to emerge.

By following the parallel with the field of Ecology, new \emph{niches} (dubbed here \emph{functions}) emerge and new companies quickly move in to seize the opportunity.
We operationalise the concept of online functions using the Crunchbase functional categories~\cite{Crunchbase2020}.
Crunchbase (\url{www.crunchbase.com}) is an index of companies, for whom a series of indicators are recorded, such as its location, number of employees, the funding rounds and the amount of money raised.
Crunchbase classifies companies using one or more labels from a taxonomy that records 744 categories, which are intended to correspond to specific market segments.
We study twelve such categories (or functions): Social network, Search, General retail, Filesharing, Music streaming, Movies \& TV, Ride sharing, Accommodation, Action cameras, Ephemeral messaging and Dating Apps for mobile.
We find that the Crunchbase categories are very broad, encompassing companies whose main business does not relate to the function (e.g. `Mattermost' is also listed as `File Sharing', when its main function obviously is `Messaging').

We perform a second pass of selection.
For each function, we study the top companies (based on the total volume of links in Reddit) and we manually select a `champion' which aligns most closely with the function (such as Uber, Spotify, AirBnb or Dropbox).
Next, we identify their top three `rivals' as of January 2017 --- using Rivalfox (now closed).
This results in a curated list containing twelve functions, and the four main rivals in each function.
For example, the function `Ride sharing' contains Uber, Lyft, Hailo and Sidecar (see the Supporting Information \nameref{S5_Table} for the complete list of functions and rivals.).
Finally, we record the date of the first link towards a company in that function --- i.e. the date that the function emerged --- and the total number of links towards companies in that function -- the total attention towards the function.

%\paragraph{Survival rates in temporal cohorts.}
\paragraph{}
We also study the increasing online competition for different temporal cohorts based on their survival rate.
We group online companies based on the year when they are first linked in Reddit, and we build 11 temporal cohorts (one for each of the years 2006 to 2016).
We follow the companies in each temporal cohort as they age, and we keep track of how many of them are still active --- i.e., have at least one link during a one-month observation period.
Note that the survival rate can increase, as some domains can remain dormant and not be linked during one or several months.
In an equal opportunity environment, the survival rate at equal ages should be similar.
However, as activity on the web grows, we expect competition to become more intense as the number of key functions having reached the maturity phase grows. 
As a result, we expect the survival of new domains in the webspace to be lower. 

%!TEX root = ../main.tex
%
% Results and Discussion can be combined.
\section*{Results and Discussion}
\label{sec:results}

We first analyse the dynamics of online attention and the observed reduction of online diversity.
Next, we study the growth of online functions and, finally, the dynamics of temporal cohorts.
Our analysis of online attention towards companies on two large social media websites reveals a number of consistent trends. 

Since online attention is a proxy for global users attention to online services and platforms such as Youtube, Etsy and Amazon, but also to offline businesses and brands such as Disney, Tiffany \& Co and Walmart, our results could potentially shed light upon the broader economic trends in the era of the web. 

In one select case study, we show that the volume of online attention to the company Tesla is predictive of its enterprise value four to twelve months in advance.
%\TODO{MAR}{Link online attention to Tesla and enterprise value}

\subsection*{Dynamics of online attention in Social Media} 

\begin{figure}[htbp]
	\centering
	\includegraphics[width=0.9\linewidth]{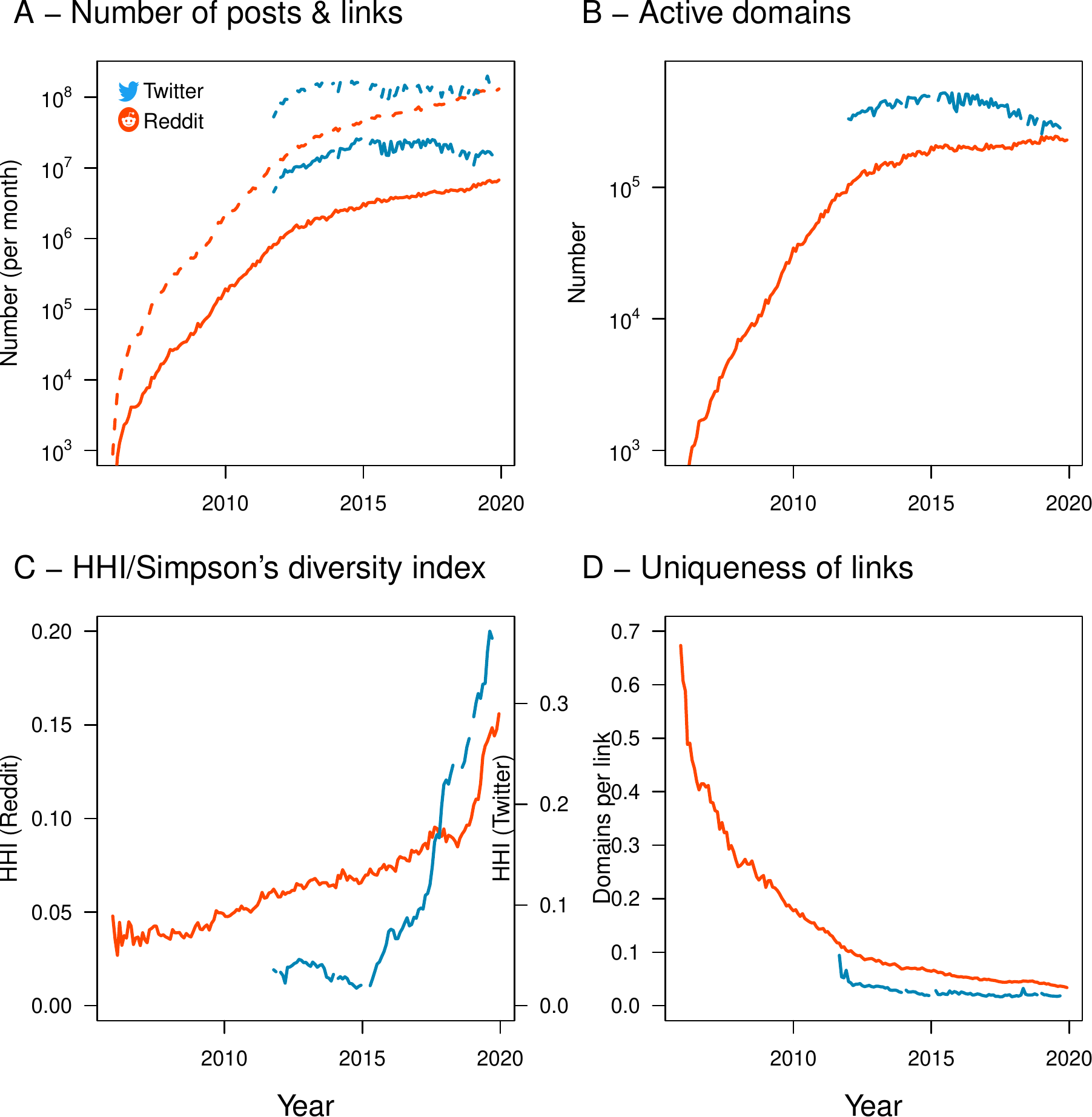}
	\vskip2em
	\caption{
		\textbf{Dynamics of activity on online platforms, as indicated via posts in social media platforms.} 
		\textbf{(A)} Growth in the number of posts (dotted) and links within posts (solid) in both Reddit and Twitter over time. 
		\textbf{(B)} The number of distinct active domains appearing within social-media links to has also grown. 
		\textbf{(C-D)} An increase the HHI index (C) and a decrease in link originality (D) for domains within links indicates that, despite the growth in total activity (A-B) diversity in online activity is in long term decline (see Sec.~\nameref{sec:materials-methods}).		
	}
	\label{fig:fig1}
\end{figure}

During the last decade, the activity on Twitter and Reddit has increased exponentially (seen in \cref{fig:fig1}A, notice the log y-axis).
% shows the number of posts and links during periods of one month for each platform.
At the same time, the total number of active domains on Reddit --- i.e. domains that have been linked at least once during a one-month period --- has increased at least two orders of magnitude from 1000 in 2006 to over 10,000 in 2020.
%links per month today. %% MAR: let's make our paper future-proof
The number of distinct and active domains linked to on Twitter is much higher and almost doubling from 246,000 in 2011 to 447,000 links per month (\cref{fig:fig1}B). 
%
%Together, these patterns highlight the flourishing and growth of activity on the web, but they give no indication towards the relative amount of attention that individual online companies receive, and more precisely how competitors fare with respect to each other.
%
%\paragraph{The exponential growth of the online environment.}
%We find that for both Reddit and Twitter, the total number of posts and web links (\cref{fig:fig1}A), and active domains (\cref{fig:fig1}B) increased exponentially across the study period.
Despite these results being consistent with the growth patterns of the web as a whole, we observe at the same time a long term decline in the diversity of services that makes up the online activity.
We measure in \cref{fig:fig1}C the Herfindahl-Hirschman Index (HHI) and in \cref{fig:fig1}D the link uniqueness (see Sec. Materials and Methods for more details).
Both figures convey the same message: in both Reddit and Twitter we observe an increasing attention concentration over time indicating that an ever increasing proportion of users attention that is focused on a smaller and smaller percentage of popular domains.
%Namely, we measure the market concentration of all domains using the Herfindahl-Hirschman Index (HHI) (see Sec. Materials and Methods for more details).}
%\verify{\cref{fig:fig1}C shows the HHI over time for active domains, for both Twitter and Reddit.}

\begin{figure}[htbp]
	\centering
	\includegraphics[width=0.9\linewidth]{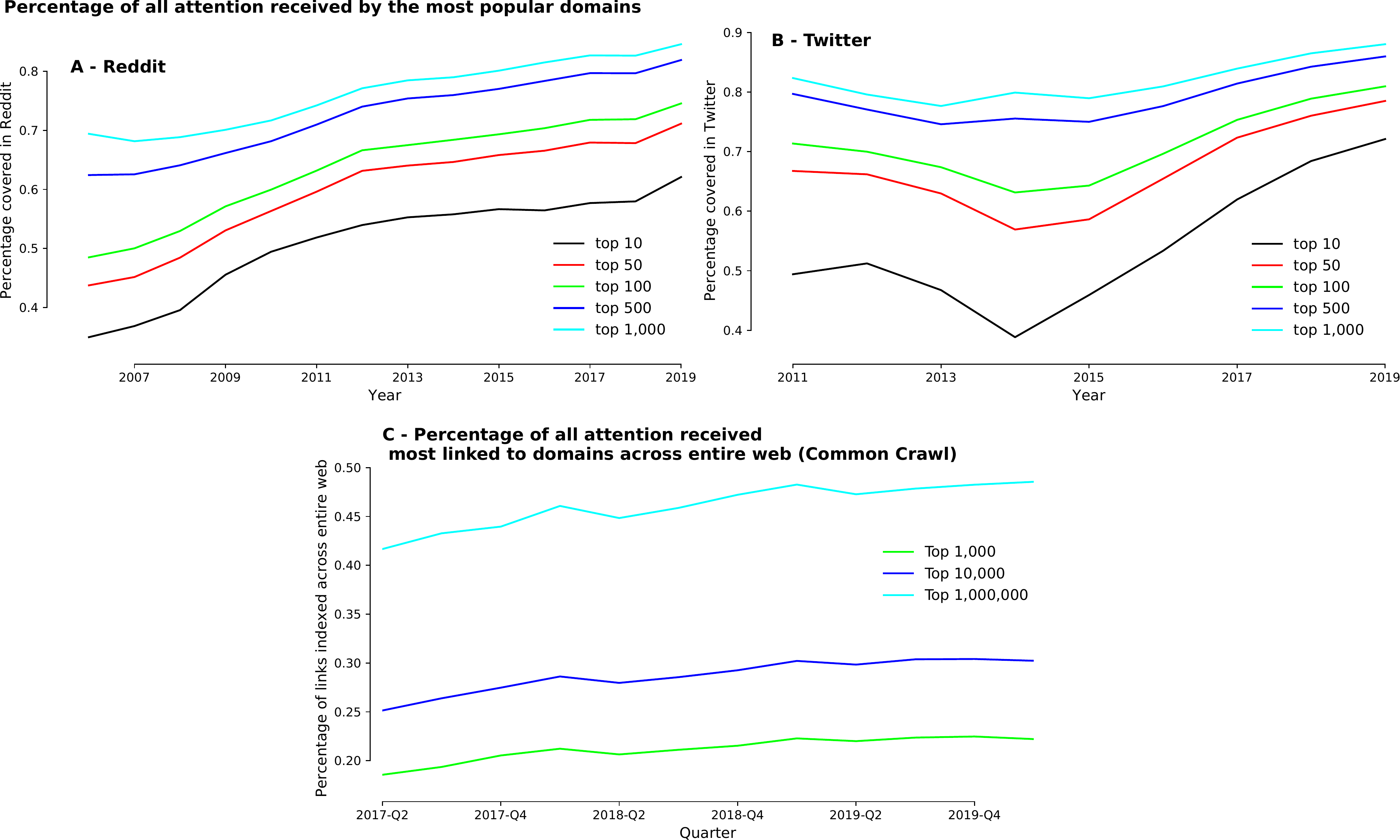}
	\vskip2em
	\caption{
		\textbf{Top websites attract growing percentages of overall links across the entire web.} 
		\textbf{(A)(B)} Percentage of all links associated with the most popular 10, 50, 100, 500 and 1,000 domains in Reddit (A) and Twitter (B). The top 10 most popular domains in Reddit received around 35\% of all links in 2006, which grew to 60\% in 2019.
		In Twitter, the top 10 domains grep from 50\% (2011) to 70\% (2019).
		\textbf{(C)} The sum of PageRank values for the top 1,000, 10,000 and 1 Million domains respectively within the Common Crawl corpus. 
	}
	\label{fig:fig2}
\end{figure}

\paragraph{Measuring the concentration of online attention.}
We can detect the temporal increase of dominance by tracking the change over time in the percentage of attention captured by the most popular domains. 
%Given the tremendous growth in online activity in our two datasets (and on the Web), we turn towards measuring how this attention is distributed among the domains corresponding to online companies.
%observe a steady decline in online diversity and an ever increasing proportion of users attention that is focused on a smaller and smaller percentage of popular domains. 
We plot in \cref{fig:fig2} the percentage of total attention received by the \emph{top n} domains, with \emph{n} between 10 and 1,000 for Reddit and Twitter.
For Reddit, \cref{fig:fig2}A shows that the top 10 most popular domains received around 35\% of all attention in 2006, which grew to about 60\% in 2019.
The percentage of attention to the top 1,000 domains (out of the total of more than 3 million on Reddit) is above 80\%.
For Twitter (\cref{fig:fig2}B), the concentration is even more pronounced, with the top 10 domains commanding about 50\% of all attention in 2011 and more than 70\% in 2019, and the top 1,000 reaching between 80\% and 90\% of all attention.
Overall, these results indicate that online media attention is very concentrated on a handful of domains, and getting increasingly concentrated over time.
Noticeably, Twitter saw a period of reduction of concentration around 2014 which was reversed towards the end of the dataset timeline.

In search for independent evidence of this concentration, we study the concentration of online attention of webpages using Common Crawl by investigating the changes in the market share of links to the top domains over time.
\cref{fig:fig2}C shows the total market share of links -- defined as the sum of their PageRank -- to the \emph{top n} most popular domaine (top 1,000, top 10,000 and top 1 million).
Similar to Reddit and Twitter, we can see that each of these grows consistently over time, clearly illustrating a growing concentration of links across the entire web among the most centrally-linked and dominant websites.
We also observe that top 1 million domains has now grown to represent almost half of the market for links in 2020. 
Even though the spread of total market share between top 10,000 and 1 Million is relatively large (around 17.4\%), the spread between top 1,000 and 10,000 is only about 7.5\%, also indicating that the top 1,000 domains occupy almost the half of the attention of the top 1 Million.

\begin{figure}[htbp]
	\centering
	\includegraphics[width=0.95\linewidth]{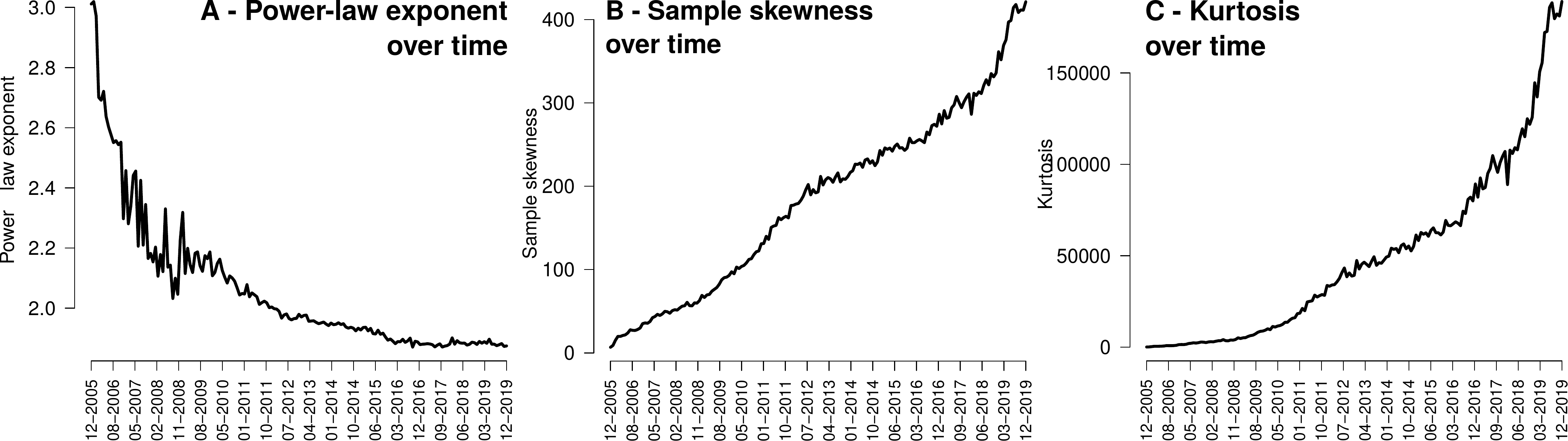}
	\vskip2em
	\caption{
		\textbf{The distribution of user attention (i.e. number of links per active domain) in Reddit, over domains, is getting increasingly more skewed over time towards the popular domains: \textit{the rich are getting richer}.} 
%		We count how many times each domain was linked during each month-long time frame, and we compute three statistics on active domains (domains that were linked at least once): 
		\textbf{(A)} we fit a power-law distribution and we plot the exponent of the power-law distribution over time; it shows an decreasing trend over time (i.e. higher inequality). 
		\textbf{(B)} and \textbf{(C)} we compute the sample skewness and kurtosis respectively; both show a upwards trend over time.
	}
	\label{fig:fig3}
\end{figure}

Next, we measure the concentration of attention from the point of view of the skewness of the attention distribution. 
We compute the attention towards domains for each one month time interval, and we measure the skewness and the kurtosis for each of these distribution. 
These are both widely used measures of distribution. A positive skewness value indicates that the tail on the right side is longer or fatter than on the left side and high kurtosis values are the result of infrequent extreme deviations (or outliers), as opposed to frequent modestly sized deviations.
Large positive values for both measures indicate a highly skewed distribution (long-tail), the larger the more skewed.
\cref{fig:fig3}B and C illustrate the skewness and the kurtosis for Reddit.
Both measures show increasingly higher values with time, indicating that attention is getting more and more dominated by several (few) domains.
 
%Removed at reviewers request We also measured the market concentration of all domains in the each dataset using the Herfindahl-Hirschman Index (HHI). 
%For both Reddit and Twitter, \cref{fig:fig1}C shows a clear ongoing increase in HHI, 
%indicating that the link distribution is increasingly concentrated, and that both platforms are getting increasingly dominated by a handful of companies.
%We note the HHI of all online services which more than doubled during the observation period indicating a smaller number of domains now online enjoy a larger relative share of the total attention as measured by their share of links in these major social media platform.  
% We also find the \emph{link originality} to be in long-term decline for both Reddit and Twitter (\cref{fig:fig1}D). 
% In 2006, over half all links on Reddit resolved to differing domains and it has steadily declined over the ensuing decade where now less than one in ten links resolve to different domains. 

\begin{figure}[htbp]
	\centering
	\includegraphics[width=0.99\linewidth]{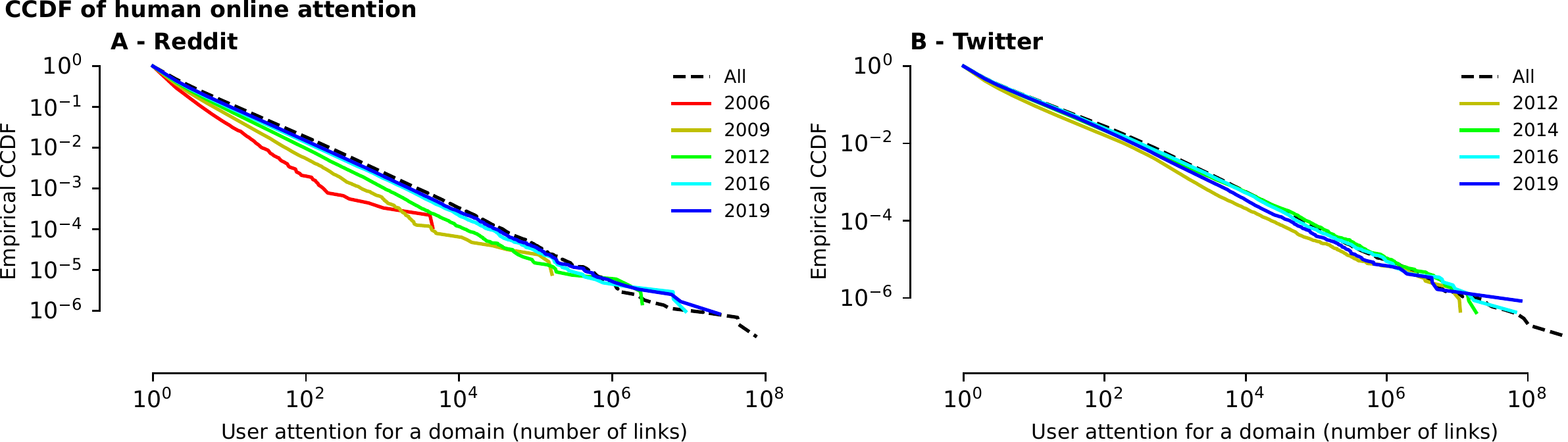}
	\vskip2em
	\caption{
		\textbf{Cumulative attention in Social Media (Reddit and Twitter) follow a rich-get-richer distribution (i.e. long-tail), which get even more skewed over time.} 
		The log-log plot of the empirical Complementary Cumulative Distribution Function (CCDF) of the number of links associated with domains in both Reddit \textbf{(A)} and Twitter \textbf{(B)} appears almost linear -- a sign of a long-tail distribution. 
%		\textbf{(B)} The log-log plot of the empirical CCDF of the number of links associated with domains in Twitter appears almost linear -- a sign of a long-tail distribution. 
		Each solid line shows the distribution of links posted during a particular year, while the dashed line shows the distribution for the entire dataset. 
% 		The distribution lines associated with later years are shifted right-upwards, since Reddit grows as a whole. 
	}
	\label{fig:fig4}
\end{figure}

\paragraph{User attention to online domains is long-tail distributed.}
The above observations indicate that online attention is long-tail distributed, with an increasingly longer tail over time.
Such distributions are also known as ``rich-get-richer'', because most of the total attention is captured by a small number of online domains.
To test this hypothesis,
%We measure user attention towards an online domain as the number of times users link the given domain in their posts and comments. 
we plot in \cref{fig:fig4} the log-log plots of the empirical Complementary Cumulative Distribution Function (CCDF) of the number of links for domains over time, in Reddit and Twitter.
Visibly, the CCDF appears linear, which is indicative of the long-tail distribution of attention. 
We also analyse the distribution of attention with given time periods, by counting only the links that were posted in particular years (here 2006, 2009, 2012, 2016, and 2019 for Reddit and 2012, 2014, 2016 and 2019 for Twitter). 
The attention pattern referred to the overall period and to any of the selected years and the distribution lines associated with later years are shifted right-upwards, since Reddit grows as a whole. Unlike Reddit, Twitter does not shift, as Twitter as a whole peaks in size in 2016, and decreases ever since (shown partially in \cref{fig:fig1}A, and in the zoom-in in \nameref{S2_Fig} in the Supporting information).

We search for a theoretical long-tail distribution that fits best the distribution of online attention.
We fit the data to several theoretical long-tail distributions (including power-law, exponential and log-normal)~\cite{Clauset2009}, and we observe that the power-law distribution has the tightest fit measured using log-likelihood ratios over the studied periods of time (see the detailed analysis in the Supporting Information).
Next, we set to detect the concentration of online attention via the fitting of power-law distributions for each of the monthly attention of domains.
Let X be a random variable, the attention received by a domain during a given time period. The probability function $P(X = x)$ decays as a power-law with parameter $\alpha$, which controls the speed of the decay. 
This can be interpreted as follows: 
for high $\alpha$ values, the probability of observing of a domain having received more attention than $x$ (the CCDF $P(X \geq x)$) decreases faster than for lower values of $\alpha$. 
Consequently, for large $\alpha$ it is less likely to have very popular domains, i.e. less dominance and more diversity. 
Conversely, lower $\alpha$ is indicative of more dominance and lower diversity. 
\cref{fig:fig3}A plots the evolution over time of the fitted value of $\alpha$.
Visibly,
%The same conclusion is further reinforced by \cref{fig:fig3}A, where 
the $\alpha$ exponent is decreasing over time, which implies that the existence of massive giants is increasingly likely.

In conclusion, the above observations show that despite the fact we observe dramatic growth of the overall web and in major social platforms such as Reddit and Twitter, we also see at the same time a long term decline in the diversity of services that makes up this online activity, with increasingly fewer players controlling increasingly larger shares of the market.

%\verify{
%Starting from the observations that the cumulative online attention that companies receive online follows a ``rich-get-richer'' distribution (see \hl{ref?}), we show that originality too follows an asymmetric distribution (\hl{ref?}). 
%Confirming the HHI analysis, \cref{fig:fig1}D shows that the link originality decreases over time in both dataset, implying an increasingly skewed distribution of attention.}

\subsection*{Growth of functionally diverse services} 

Business innovation is often a result of the emergence of new enabling platform technologies --- such as geolocation, security and broadband combined with a critical mass of users with access to this technology. 
For example, online \emph{General retail} has been enabled by the roll-out of secure online payments; 
online \emph{Video} with more broadband access and others like \emph{Ride sharing} with the widespread adoption of smartphones.

Many services themselves enable others to flourish too. 
For example before smartphones, the advent and widespread adoption of \emph{Webmail} services and public internet cafes enabled large numbers of travellers to check email and adjust and rebook travel plans while on the move thus advancing the growth of online \emph{Accommodation} services.

Others are an effective and orchestrated combination of a range of other technologies. 
\emph{Ride sharing}, for example, has mixed geolocation, secure payment and instant messaging to create a global alternative to the taxi industry.
These foundation online technologies and services form the basis to enable more and more complex new services to be offered online, and their networked nature means the widespread adoption and diffusion of new services is increasingly rapid.

In this section we analyse the dynamics of the online attention economy in waves of innovation.
%Moved from materials and methods as requested by reviewer 
We show the temporal emergence of online functions and we study the dynamics of the survival rates for online companies.

%\verify{\paragraph{Four waves of online innovation.}
While the web started as an information media, with the addition of security, mobility and broadband access it has quickly evolved as a marketplace for services. 
Waves of business and services innovation have followed each wave of infrastructure innovation and we identify four such waves.

\textbf{In the first wave}, simple text based information services emerged such as an online reference about the cast and crew of almost all movies ever made (IMDB 1990), a platform for online diaries (Blogger 1993) and a service for online classifieds (craigslist 1995). 
\textbf{In the second wave}, with addition of online cryptography and security, a raft of new commercial services emerged such as Online Retailers (Amazon 1994); Online Classified Auctions (eBay 1995) and Online Payment (PayPal 1998).
\textbf{In the third wave}, the ability for large number of internet users to be connected to high-capacity broadband led to more innovation in media and communications services such as Online Telecommunications (Skype 2003), Online DIY Video (YouTube 2005), and Online Movies (Netflix Streaming 2007). 
Finally, \textbf{in the forth wave}, the advent of online mobility created services to track movement online (fitbit 2007); create new marketplaces for accommodation (Airbnb 2008) and ridesharing (Uber 2009).

\begin{figure}[htbp]
	\centering
	\includegraphics[width=0.99\linewidth]{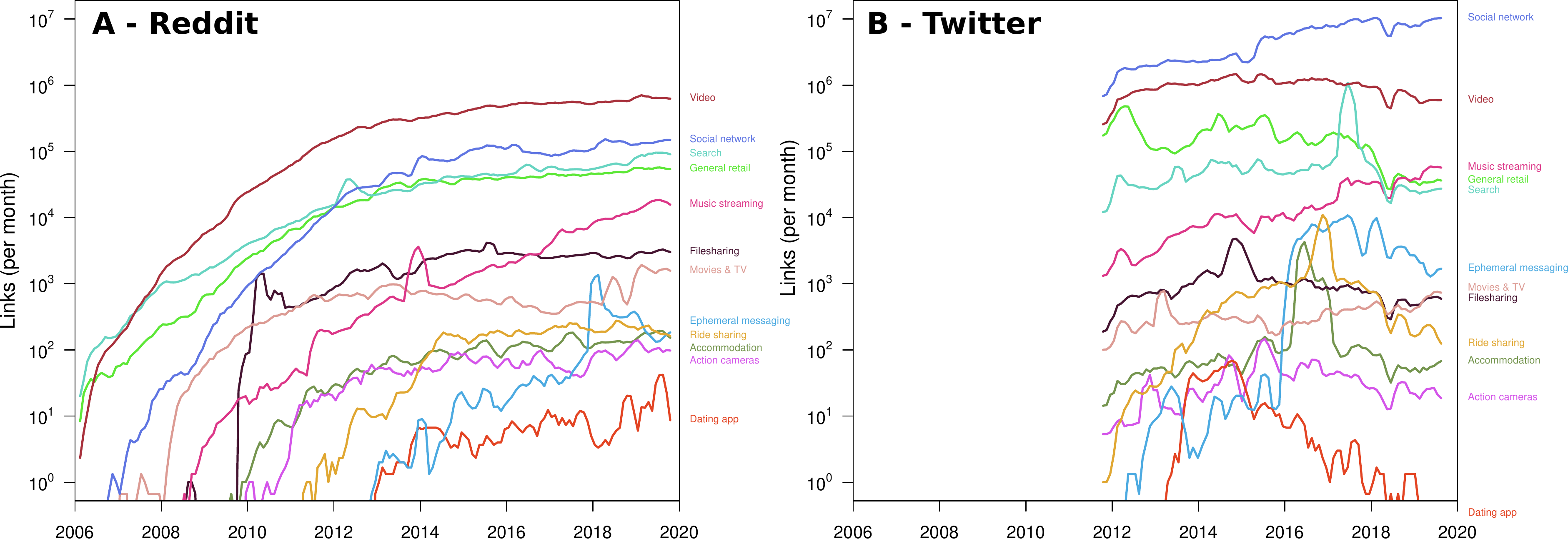}
	\vskip2em
	\caption{
		\textbf{Growth of functional diversity on the Reddit (A) and Twitter (B) platforms.} 
		The lines shows the number of links to different functions as indicated by the leading organisations and their top-three rivals (see \textit{Sec.~\nameref{subsec:attention-economy}} for details).
	}
	\label{fig:fig5}
\end{figure}

We postulate that the above-defined waves of technology innovation lay the foundations for new types of services or functions to emerge. These new functions compete not only with other companies on what they are selling but in the ways of delivering products and services and thus are often disruptive to existing players as they circumvent existing approaches to market.  New functions also provide new avenues for emerging companies to enter the market and often a key part of the competitive advantage of startups. 
Whether delivered by startups or established firms who reinventing themselves via digital transformation, new technology-enabled functionally diverse services forms the basis for much of the disruptive innovation we have seen over the past two decades.

Our measurements provide empirical proof for this hypothesis, as they show that different functions appear at different times, and they follow similar patterns of increasing activity.
\cref{fig:fig5}A shows that in Reddit new functions continuously emerge and the attention towards each of them grows consistently, exponentially at first (note the log scale of the y-axis) and at more moderate rates as the functions mature.
%Note that the y-axis is in log scale, which implies an exponential growth.
%rapidly at first then plateaus but continues to accumulate.  
\cref{fig:fig5}B provides comparable conclusions for Twitter, but over a shorter timeframe. Note the relative decline in links to some functions on the largely mobile user base of Twitter such as dating, accommodation and filesharing - possibly due to the widespread uptake of mobile apps in these areas.

\begin{figure}[htbp]
	\centering
	\includegraphics[width=0.99\linewidth]{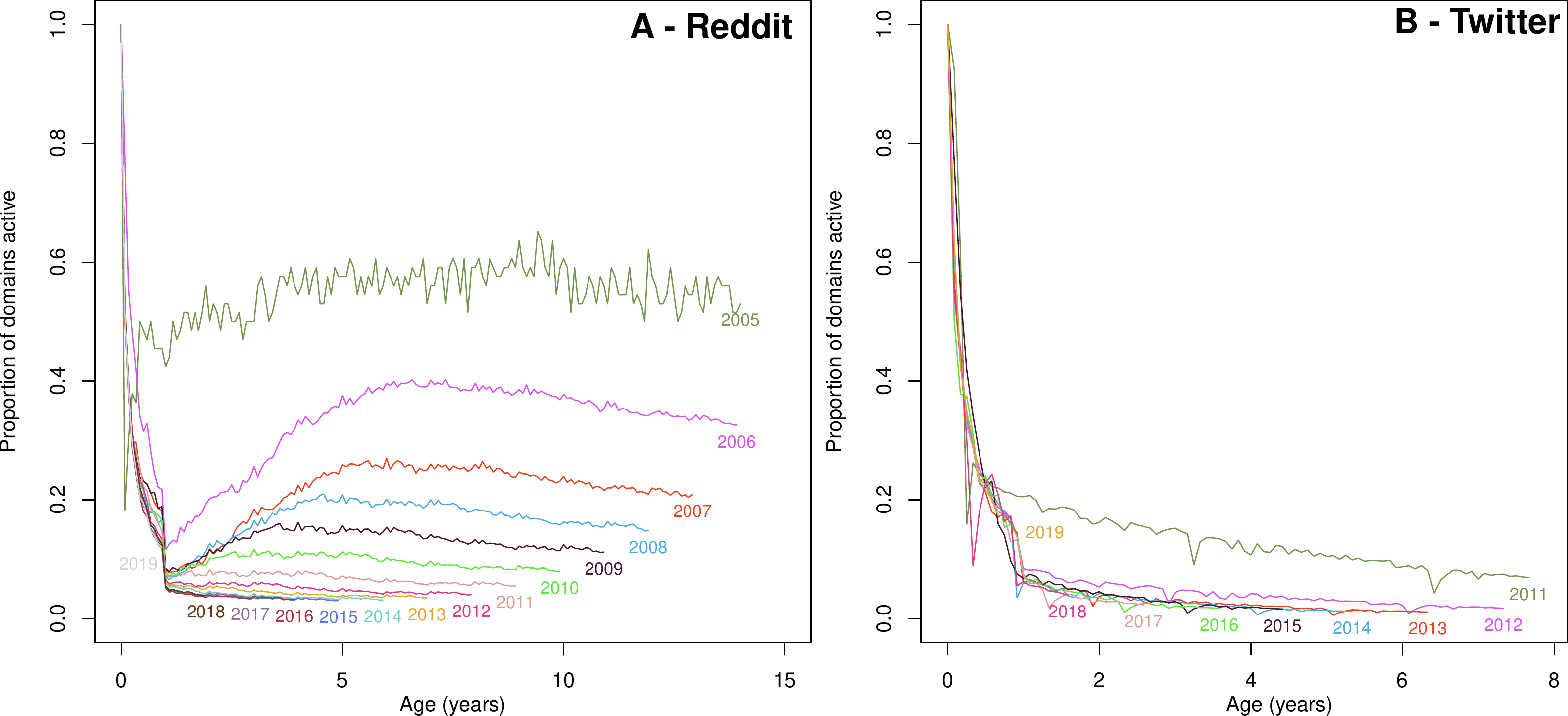}
	\vskip2em	
	\caption{
		\textbf{Survival rates of newborn online services are in decline} in both Reddit \textbf{(A)} and Twitter \textbf{(A)}. 
		Lines show survival rates for the cohort of all new domains sorted by year of first appearance. 
		The percentage still active 5 years after their first appearance from that birth year is in steady decline. 
		Survival may increase after year 0%, as new domains often fall dormant for a period after first entry.
	}
	\label{fig:fig6}
\end{figure}

\subsection*{Survival rates for newborn online services are in decline}

For a company, the success in securing online attention correlates with business success.
Conversely, the lack of online attention can signal
%As links on major social media platforms to domains are a sign of activity - their absence can equally be used as a signal of now 
a decline in customer demand or defunct services.
In our analysis, a domain's `birth' occurs with the first link to have ever been posted towards the domain, while its `death' occurs with the last link that points to it --- obviously there is an edge effect at the end of our dataset in 2019, which we account for by not over-interpreting the results for the last cohort of 2018.

In \cref{fig:fig6}, we group domains in temporal cohorts based on the year of their birth (see Sec. ``\nameref{subsec:attention-economy}''). 
Each line in \cref{fig:fig6} corresponds to a temporal cohort, and
%By looking at domains by birth cohorts based on the year they first appear with links, 
we observe that survival rates of newborn online services are in decline. 
For example, the percentage of domains still active 5 years after their birth year has declined from just under 40\% for the 2006 cohort to less than 5\% for the 2019 cohort.
That is to say, a smaller proportion of newborn domains survive to older ages in later cohorts, which indicates that the competitive environment for young firms is becoming more hostile over time.

Consistent with previous research that shows that growing digitisation of industries stymies new entrants\cite{wang2015role}, our analysis reveals \emph{mortality rates} of new entrants online are on the rise.
This indicates the environment for new players online is becoming increasingly difficult.
In the way pine trees sterilise the ground under their branches to prevent other trees competing with them, once they are established dominant players online crowd out competitors in their functional niche.

Let's take search as an illustrative example.
Google was founded in 1998 and fought off many early rivals such as AltaVista, Yahoo and Hotbot to the crown of the world's search engine of choice. 
Another serious competitor to Google, Cuil, emerged late on the scene some eight years later in 2006 but by this stage Google's market share was clearly dominant.
Despite attracting over \textdollar 30 million in investment from leading Venture Capital firms and indexing more websites than  Google, Cuil was unable to make the slightest dent on Googles market share --- achieving only 0.2 percent of worldwide search engine users in July 2008 and the service was shut down in 2010. 
Consequently, Google has gone on to dominate search in the English-speaking world with over 90 percent market share in the past decade~\cite{Statscounter2020}.
%Source: \url{https://gs.statcounter.com/search-engine-market-share\#monthly-200901-202001}

\subsection*{Linking online attention to enterprise value}
\begin{figure}[htbp]
	\centering
	\includegraphics[width=0.95\linewidth]{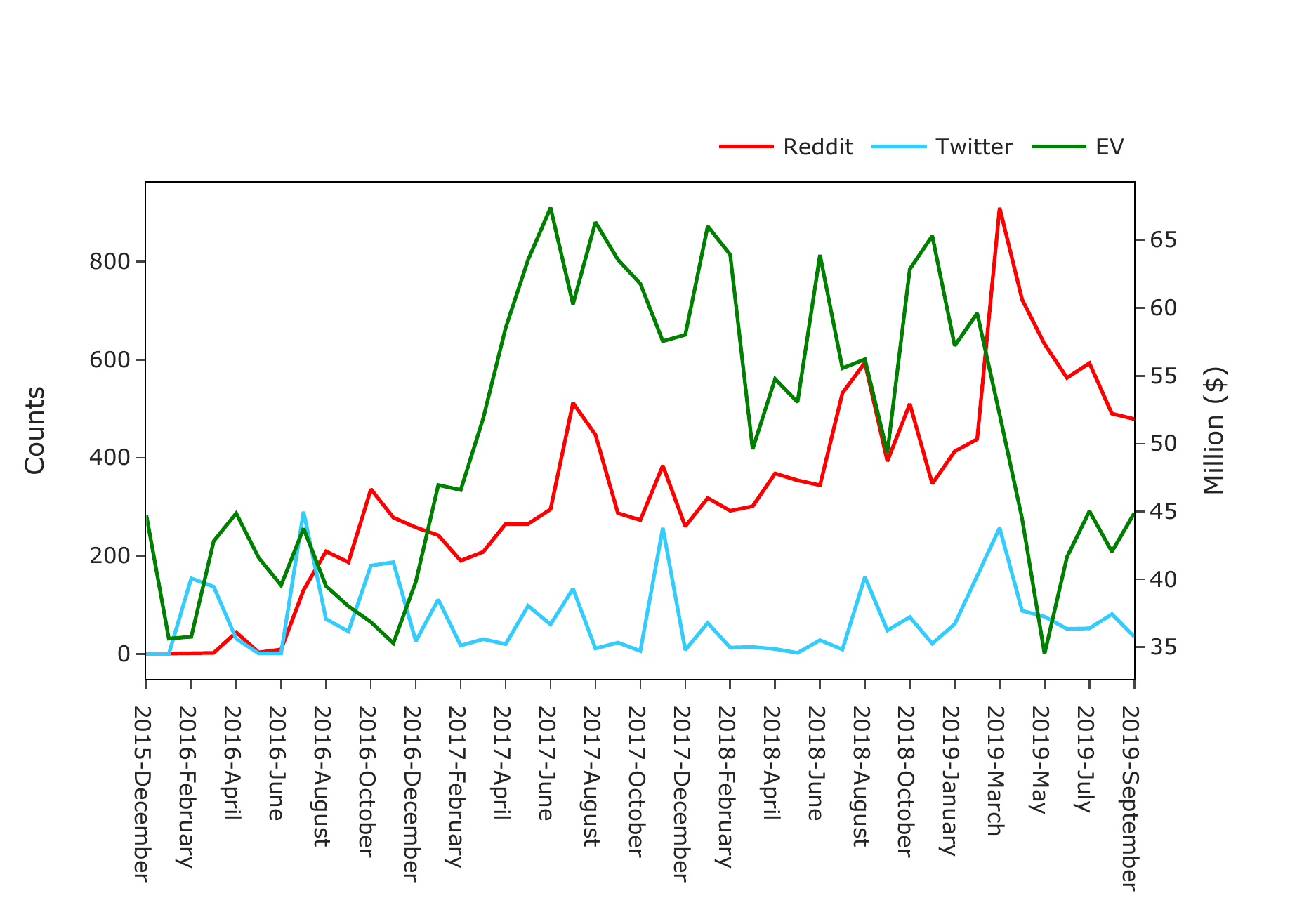}
	\vskip2em
	\caption{
		\textbf{Tesla, Inc (NASDAQ: TSLA)}: Time-series trends of enterprise value (EV, \textcolor{ForestGreen}{green line}, shown on right y-axis) vs online attention (Reddit, \textcolor{red}{red line}, and Twitter \textcolor{blue}{blue line}, shown on the left y-axis).
	}
	\label{fig:fig7}
\end{figure}

Here, we conduct a case study indicative of the link between the share of attention received on online social media and offline financial performances of companies.
We build the Enterprise Value (EV) time series of electric vehicle manufacturer Tesla, Inc (NASDAQ: TSLA)  between December 2015 and September 2019 and we present it in \cref{fig:fig7} together with the attention series in Reddit and Twitter.
Using the methodology outlined in \nameref{sec:materials-methods} (and further detailed in the Supplementary Information), we show that both the series for Twitter and Reddit are Granger-causal for the EV (see \nameref{sec:materials-methods}).
In other words, this result indicates that social media attention is a leading indicator for trends in growth in investors re-evaluation of Tesla's value -- i.e. the past online attention is predictive for the changes in enterprise value in the future.
Furthermore, we can identify the `optimal' lag between social attention and EV to being four months for Twitter and twelve months for Reddit.

%By linking time-series financial market data we have observed that the  and link counts from our social media data sets in Reddit and Twitter are related.  \cref{fig:fig7} shows the link counts and enterprise value of.
%Using statistical analysis of this data, we found the growth in links on social media are predictive of a growth in enterprise value for Tesla. We found link counts in both Twitter and Reddit to be Granger causally related to the growth in enterprise value.
%More details of the analysis and results are included in the supporting information.
%explore whether posts on news-centric Twitter may indeed be a leading indicator for trends in growth in investors re-evaluation of Tesla's value by looking at tests for various lags between counts of links to Tesla.com Twitter and subsequent changes in its Enterprise Value. 

This final result is significant, as it indicates a pathway of translating the increasing concentration of online attention to a handful of companies with online presence to an increasing dominance of these companies in the offline world given their boosted financial position.
Of course, a wide array of factors determines the success of companies in the offline world, but we posit that the concentration of online attention might prove to be the one extra boost that can make the difference between growth and failure down the line.

%!TEX root = ../main.tex
%
\section*{Conclusion}
This study addresses an apparent paradox:
the web is a source of continual innovation\cite{sawhney2005collaborating}, and yet it appears increasingly dominated by a small number of dominant players\cite{haucap2014google}. 
This research tackles this paradox by using large-scale longitudinal data sets from social media to measure the distribution of attention across the whole online economy over more than a decade from 2006 until 2017. 
Here, we use outbound weblinks towards distinguishable web domains as a proxy for the market for online attention.
As this data collection captures longitudinal trends relating to a universe of all potential websites and services, it serves as a valuable index of broader economic trends, dynamics, and patterns emerging online. 

In this work, we provided evidence consistent with a link between increasing online attention on social media and the emergence (and growing) dominance of a small number of players. However, the question remains open concerning the real causes of dominance: is online attention part of them or just an early indicator? While it is impossible to infer causal relations from large observational data, such as those used here, our results are consistent with a putative link.
% , the question remains whether is social media attention a key cause of this dominance or correlated to it.

The development of the web has been steady, and it came in functional waves, each of which has been predicated by the emergence of foundation platform technologies--- such as secure encryption, enabling e-commerce; ubiquitous broadband, enabling the emergence of streaming video and mobility, enabling the emergence of many new functions including car sharing.
This research outlines that while new functions, services, and business models continuously emerge online, the web dynamics are such that in many mature categories of online services, one or a small number of competitors dominate.
Yet, as new web technologies continue to be developed, this enables more unexplored functional niches to emerge and for the cycle to repeat\cite{McCarthy2015}. 
Over time, this process leads to long-term declines in the overall competition, diversity, and decreasing survival rates for new entrants.

%% MAR: this is repetitive
%The web is now synonymous with the birth and growth of new products, services and business models. 
%Yet in many mature categories of online services, one or a small number of competitors dominate. 

The world's largest companies are now those that run global online platforms: Apple, Facebook, Google, and Amazon in the west and their counterparts Alibaba, Baidu, and Tencent in China. 
There is a growing public interest in the nature and extent of dominance on the web and web giants' influence on economics, popular culture, and even politics. 
This paper extends understanding of the nature and scope of the web's network effects on the evolution of businesses today. 
This work also opens the door to further research that uses digital footprints of organisations \emph{en masse} as a basis for analysis of the behavioural economics and competitive dynamics of markets online. There is room here too for further work in simulation extending previous work done in synthetic market experimentation and prediction.~\cite{lera2020prediction,Salganik2006}.

While Twitter and Reddit might not be representative for the web activity worldwide, they serve as a good representative sample of activity of most activity online in the English speaking web.
Future work could compliment the analysis with equivalent datasets of online ecosystems in China (\url{www.weibo.com}, \url{www.weechat.org}), Russia (\url{www.vk.com}, \url{www.yandex.com}) and others.

As the web is an integral part of every industry, this research's scope extends beyond technology firms~\cite{Albert1999,Huberman1999,Gandhi2016}. 
Although the world's largest technology giants are now also the largest companies globally, the technology sector itself still represents a small but growing subset of the overall economy. 
However, the reach and impact of digitisation, and online information and services has been shown to impact over 98\% of the entire economy~\cite{manyika2015digital}. 
And while the data used in this study (from links in billions of online posts) reflects only the online activity, we posit that the patterns identified here represent the trends across the entire economy.

\section*{Acknowledgments}
We thank G. Brewer (Builtwith.com) and J. Baumgartner (pushshift.io) for their assistance with the initial exploratory analysis. 
We also thanks S. Nagel (Commoncrawl.org) and R. Viscomi (google.com) for their kind advice and help using large-scale public open web indexes Commoncrawl and HTTP Archive. 
This research was supported by UNSW Australia, UTS, and ANU. 
Daniel Falster was supported by the Australian Research Council (FT160100113), and Marian-Andrei Rizoiu was partially supported by Facebook Research under the Content Policy Research Initiative grants and by the Commonwealth of Australia (represented by the Defence Science and Technology Group). 
The funders had no role in study design, data collection and analysis, decision to publish, or preparation of the manuscript.

% \nolinenumbers

% Either type in your references using
% \begin{thebibliography}{}
% \bibitem{}
% Text
% \end{thebibliography}
%
% or
%
% Compile your BiBTeX database using our plos2015.bst
% style file and paste the contents of your .bbl file
% here. See http://journals.plos.org/plosone/s/latex for 
% step-by-step instructions.
% 
%\bibliographystyle{plos2015}
\bibliography{refs}

\begin{thebibliography}{10}

\bibitem{friedman2005world}
Friedman TL.
\newblock The world is flat: A brief history of the twenty-first century.
\newblock Macmillan; 2005.

\bibitem{levine2009cluetrain}
Levine R, Locke C, Searls D, Weinberger D.
\newblock The cluetrain manifesto.
\newblock Basic books; 2009.

\bibitem{da2011search}
Da Z, Engelberg J, Gao P.
\newblock In search of attention.
\newblock The Journal of Finance. 2011;66(5):1461--1499.

\bibitem{xiang2010role}
Xiang Z, Gretzel U.
\newblock Role of social media in online travel information search.
\newblock Tourism management. 2010;31(2):179--188.

\bibitem{McCarthy2015}
McCarthy PX.
\newblock Online Gravity: The Unseen Force Driving the Way You Live, Earn, and
  Learn.
\newblock Simon and Schuster; 2015.

\bibitem{Scheffer2017}
Scheffer M, van Bavel B, van~de Leemput IA, van Nes EH.
\newblock {Inequality in nature and society.}
\newblock Proceedings of the National Academy of Sciences of the United States
  of America. 2017;114(50):13154--13157.
\newblock doi:{10.1073/pnas.1706412114}.

\bibitem{shapiro1998information}
Shapiro C, Varian HR.
\newblock Information rules: a strategic guide to the network economy.
\newblock Harvard Business Press; 1998.

\bibitem{wang2015role}
Wang F, Zhang XPS.
\newblock The role of the Internet in changing industry competition.
\newblock Information \& Management. 2015;52(1):71--81.

\bibitem{Danescu-Niculescu-Mizil2010}
Danescu-Niculescu-Mizil C, Broder AZ, Gabrilovich E, Josifovski V, Pang B.
\newblock {Competing for users' attention}.
\newblock In: Proceedings of the 19th international conference on World wide
  web - WWW '10. New York, New York, USA: ACM Press; 2010. p. 291.

\bibitem{Rizoiu2017}
Rizoiu MA, Xie L, Sanner S, Cebrian M, Yu H, {Van Hentenryck} P.
\newblock {Expecting to be HIP: Hawkes Intensity Processes for Social Media
  Popularity}.
\newblock In: Proceedings of the 26th International Conference on World Wide
  Web (WWW '17). New York, New York, USA: ACM Press; 2017. p. 735--744.

\bibitem{Wu2019}
Wu S, Rizoiu MA, Xie L.
\newblock {Estimating Attention Flow in Online Video Networks}.
\newblock Proceedings of the ACM on Human-Computer Interaction.
  2019;3(CSCW):1--25.
\newblock doi:{10.1145/3359285}.

\bibitem{Mishra2016}
Mishra S, Rizoiu MA, Xie L.
\newblock {Feature Driven and Point Process Approaches for Popularity
  Prediction}.
\newblock In: Proceedings of the 25th ACM International on Conference on
  Information and Knowledge Management - CIKM '16. Indianapolis, IN, USA: ACM
  Press; 2016. p. 1069--1078.

\bibitem{Zhang2019}
Zhang R, Walder C, Rizoiu MA, Xie L.
\newblock {Efficient Non-parametric Bayesian Hawkes Processes}.
\newblock In: Proceedings of the Twenty-Eighth International Joint Conference
  on Artificial Intelligence. California: International Joint Conferences on
  Artificial Intelligence Organization; 2019. p. 4299--4305.

\bibitem{jun2014study}
Jun SP, Yeom J, Son JK.
\newblock A study of the method using search traffic to analyze new technology
  adoption.
\newblock Technological Forecasting and Social Change. 2014;81:82--95.

\bibitem{d2017predictive}
D’Amuri F, Marcucci J.
\newblock The predictive power of Google searches in forecasting US
  unemployment.
\newblock International Journal of Forecasting. 2017;33(4):801--816.

\bibitem{wu2015future}
Wu L, Brynjolfsson E.
\newblock The future of prediction: How Google searches foreshadow housing
  prices and sales.
\newblock In: Economic analysis of the digital economy. University of Chicago
  Press; 2015. p. 89--118.

\bibitem{ginsberg2009detecting}
Ginsberg J, Mohebbi MH, Patel RS, Brammer L, Smolinski MS, Brilliant L.
\newblock Detecting influenza epidemics using search engine query data.
\newblock Nature. 2009;457(7232):1012.

\bibitem{mciver2014wikipedia}
McIver DJ, Brownstein JS.
\newblock Wikipedia usage estimates prevalence of influenza-like illness in the
  United States in near real-time.
\newblock PLoS computational biology. 2014;10(4):e1003581.

\bibitem{Rizoiu2016}
Rizoiu MA, Xie L, Caetano T, Cebrian M.
\newblock In: International Conference on Web Search and Data Mining (WSDM
  '16). ACM. New York, New York, USA: ACM Press; 2016. p. 215--224.

\bibitem{aramaki2011twitter}
Aramaki E, Maskawa S, Morita M.
\newblock Twitter catches the flu: detecting influenza epidemics using Twitter.
\newblock In: Proceedings of the conference on empirical methods in natural
  language processing. Association for Computational Linguistics; 2011. p.
  1568--1576.

\bibitem{kutcher2014grow}
Kutcher E, Nottebohm O, Sprague K.
\newblock Grow fast or die slow.
\newblock McKinsey \& Company (http://www mckinsey
  com/Insights/High\_Tech\_Telecoms\_Internet/Grow\_fast\_or\_die\_slow).
  2014;.

\bibitem{lera2020prediction}
Lera SC, Pentland A, Sornette D.
\newblock Prediction and prevention of disproportionally dominant agents in
  complex networks.
\newblock Proceedings of the National Academy of Sciences. 2020;.

\bibitem{gupta2004valuing}
Gupta S, Lehmann DR, Stuart JA.
\newblock Valuing customers.
\newblock Journal of marketing research. 2004;41(1):7--18.

\bibitem{bozovic2017unicorns}
Bozovic D, Sornette D, Wheatley S. Unicorns Analysis: An Estimation of
  Spotify's and Snapchat's Valuation; 2017.

\bibitem{ITU2020}
ITU UN. {ITU Statistics: Internet usage}; 2020.
\newblock Available from:
  \url{https://www.itu.int/en/ITU-D/Statistics/Pages/stat/default.aspx}.

\bibitem{BondInternet2019}
{Bond Internet Trends}.
\newblock {Daily hours spent with digital media, United States}.
\newblock Bond; 2019.
\newblock Available from: \url{https://www.bondcap.com/report/itr19/}.

\bibitem{Internet2006}
{Zooknik Internet Intelligence}. {History of gTLD domain name growth}; 2006.
\newblock Available from: \url{http://zooknic.com/Domains/counts.html}.

\bibitem{Verisign2016}
Verisign.
\newblock {The Domain Name Industry Brief}.
\newblock Verisign; 2016.
\newblock Available from:
  \url{https://www.verisign.com/assets/domain-name-report-Q42016.pdf}.

\bibitem{arthur1989competing}
Arthur WB.
\newblock Competing technologies, increasing returns, and lock-in by historical
  events.
\newblock The economic journal. 1989;99(394):116--131.

\bibitem{barabasi1999emergence}
Barab{\'a}si AL, Albert R.
\newblock Emergence of scaling in random networks.
\newblock Science. 1999;286(5439):509--512.

\bibitem{watts2002simple}
Watts DJ.
\newblock A simple model of global cascades on random networks.
\newblock Proceedings of the National Academy of Sciences.
  2002;99(9):5766--5771.

\bibitem{adamic2002zipf}
Adamic LA, Huberman BA.
\newblock Zipf's law and the Internet.
\newblock Glottometrics. 2002;3(1):143--150.

\bibitem{Waters2017Reddit}
Waters R.
\newblock {Reddit caught between user highs and ‘edgier' lows}.
\newblock Financial Times. 2107;.

\bibitem{page1999pagerank}
Page L, Brin S, Motwani R, Winograd T.
\newblock The PageRank citation ranking: Bringing order to the web.
\newblock Stanford InfoLab; 1999.

\bibitem{pop00001}
Nickell S.
\newblock Competition and corporate performance.
\newblock Journal of political economy. 1996;.

\bibitem{arrow1962economic}
Arrow K.
\newblock {Economic Welfare and the Allocation of Resources for Invention}.
\newblock In: Nelson R, editor. The Rate and Direction of Inventive Activity:
  Economic and Social Factors. Princeton University Press; 1962. p. 609--626.

\bibitem{CommonCorpusCoverage}
Corpus C. {Estimation of Representativeness of a Recent Crawl}; 2020.
\newblock Available from:
  \url{https://commoncrawl.github.io/cc-crawl-statistics/plots/tld/comparison.html}.

\bibitem{simpson_measurement_1949}
Simpson EH.
\newblock Measurement of {Diversity}.
\newblock Nature. 1949;163(4148):688.
\newblock doi:{10.1038/163688a0}.

\bibitem{hirschman1964paternity}
Hirschman AO.
\newblock The paternity of an index.
\newblock The American Economic Review. 1964;54(5):761--762.

\bibitem{RCore}
{R Core Team}. R: A Language and Environment for Statistical Computing; 2017.
\newblock Available from: \url{https://www.R-project.org/}.

\bibitem{Crunchbase2020}
Crunchbase S. {Functional categories in Crunchbase}; 2020.
\newblock Available from:
  \url{https://support.crunchbase.com/hc/en-us/articles/360009616373-What-categories-are-included-in-Crunchbase-}.

\bibitem{Clauset2009}
Clauset A, Shalizi CR, Newman MEJ.
\newblock {Power-Law Distributions in Empirical Data}.
\newblock SIAM Review. 2009;51(4):661--703.
\newblock doi:{10.1137/070710111}.

\bibitem{Statscounter2020}
Statscounter G. {Search Engine Market Share Worldwide: Jan 2009 - Jan 2020};
  2020.
\newblock Available from:
  \url{https://gs.statcounter.com/search-engine-market-share{\#}monthly-200901-202001}.

\bibitem{sawhney2005collaborating}
Sawhney M, Verona G, Prandelli E.
\newblock Collaborating to create: The Internet as a platform for customer
  engagement in product innovation.
\newblock Journal of interactive marketing. 2005;19(4):4--17.

\bibitem{haucap2014google}
Haucap J, Heimeshoff U.
\newblock Google, Facebook, Amazon, eBay: Is the Internet driving competition
  or market monopolization?
\newblock International Economics and Economic Policy. 2014;11(1-2):49--61.

\bibitem{Salganik2006}
Salganik MJ, Dodds PS, Watts DJ.
\newblock {Experimental study of inequality and unpredictability in an
  artificial cultural market.}
\newblock Science (New York, NY). 2006;311(5762):854--6.
\newblock doi:{10.1126/science.1121066}.

\bibitem{Albert1999}
Albert R, Jeong H, Barab{\'{a}}si AL.
\newblock {Diameter of the World-Wide Web}.
\newblock Nature. 1999;401(6749):130--131.
\newblock doi:{10.1038/43601}.

\bibitem{Huberman1999}
Huberman BA, Adamic LA.
\newblock {Growth dynamics of the World-Wide Web}.
\newblock Nature. 1999;401(6749):131--131.
\newblock doi:{10.1038/43604}.

\bibitem{Gandhi2016}
Gandhi P, Khanna S, Ramaswamy S.
\newblock {Which Industries Are the Most Digital (and Why)?}
\newblock Harvard business review. 2016;1.

\bibitem{manyika2015digital}
Manyika J, Ramaswamy S, Khanna S, Sarrazin H, Pinkus G, Sethupathy G, et~al.
\newblock Digital America: A tale of the haves and have-mores.
\newblock McKinsey Global Institute. 2015;.

\end{thebibliography}

\newpage
%!TEX root = ../main.tex
%
\clearpage
\appendix 
\singlespacing
 
\section*{Supporting information}

% Include only the SI item label in the paragraph heading. Use the \nameref{label} command to cite SI items in the text.

\subsection*{S1 Appendix: sampling social media data} \label{S1_Appendix}

\textbf{Twitter sampling.}
HHI calculations in Economics typically involve looking at the market concentration of the top 50 companies in an industry or sector.
With respect to the web as a whole, there are between ten and twenty major functions that have emerged including search, video, music, retail etc. 
Consequently sampling to include the concentration of the Top 1000 should correspond to covering leaders of the 10-20 most popular functions.
Also Twitter is known to be a noisy data source. Because of the open API, many robots and re-tweeting tools are in operation which results in lots of links to sites and services that are non-mainstream.
The large-scale global independent audience measurement service Quantcast\footnote{Quantcast global audience: \url{www.quantcast.com/}} was used to determine the Top-1000 websites and HHI within this sample from twitter was used to measure the concentration of promotional robots who post large volumes of links to services that are not that well visited.

\textbf{Representativeness of the Reddit dataset}
Here we report on the representativeness of our Reddit dataset. 
We used Amazon's Alexa, which ranks websites according to their traffic worldwide.
The most visited websites are not the same as the most linked to. 
There is consistent overlap with the most popular sites, with Youtube, Snapchat and the BBC being among the most visited and linked to.
However, there are other websites that are popular, but not that frequently linked to such as advertising networks, adult sites and non-English language websites. 
Conversely there are a variety of websites that attract large volumes of links, but which do not necessarily have many direct users.
Examples include automated services that enable delayed or buffered posting of social media content, reposting of content from other social media and potentially promotional robots who post large volumes of links to services that are not that well visited.
 
We verify our samples' coverage by comparing them to the most popular sites on the web. 
\nameref{S4_Fig} shows that the majority of the most visited sites on the web are included in our sample with 95 percent of the Top 100 most visited domains in the world in 2016\footnote{Source: Amazon's Alexa's Top 1 Million Most Visited Global Websites, 30th October 2016.} are represented in our sample. 
Those that are missing comprise shortened URLs (that are later expanded) and selected advertising services and adult websites that are not widely shared among peers.

\subsection*{S2 Appendix: Enterprise Financial Performance Data} \label{S2_Appendix}

A study on the financial performance of companies is beyond the scope of this paper. 
Still, in the main text and Section \nameref{S3_Appendix}, we examine the relationship between attention on social media and firm performance of one exemplar company (Tesla) in operating in a new and emerging category (electric vehicles).  
Links on social media are a significant measure of user attention, which can often be a precursor of revenue growth, a condition of financial performance.  
Previous research has illustrated that in new and emerging product categories, such as electric vehicles, competing brands' market share can be predicted by the percentage of links~\cite{jun2014study}.  
Tesla is now a pioneer in electric vehicles and is perceived as a technology leader in its own right. 
In 2020, Tesla overtook all other carmakers in terms of its enterprise value (as measured by market capitalisation) to become the world's most valuable car company\footnote{\url{https://www.forbes.com/sites/sergeiklebnikov/2020/07/01/tesla-is-now-the-worlds-most-valuable-car-company-with-a-valuation-of-208-billion/?sh=6742e4d55334}}.
As Tesla's current production volumes, revenue and earnings are significantly lower than other large global carmakers, this valuation is based largely on investors expectations of future earnings potential and growth of the company.

Preceding this rise in valuation, Telsa's brand has also attracted a more significant share of Wikipedia page visits since 2017 than its rivals (all top 10 carmakers\footnote{Top 10 carmakes: \url{https://www.interbrand.com/best-global-brands/?filter-brand-sector=automotive}}, shown in Fig.~\nameref{S5_Fig}A). 
Currently, Tesla attracts more Google searches in the US than all other rivals except Toyota, having overtaken Ferrari in 2012, GM in 2013, and VW in 2016 (shown in Fig.~\nameref{S5_Fig}B).
Our extended analysis (detailed in the following Section~\nameref{S3_Appendix}) shows that growth in the number of links in social media in both Reddit and Twitter to \url{Tesla.com} precedes an increase in the enterprise value of Tesla and that this link is Granger-causal.
% We show there to be Granger causality between social links and enterprise value.
Together, these provide indirect evidence of the connection between online attention and offline market value of companies.

% While beyond the scope of this current paper to look at detailed business performance metrics we have added, data and analysis of the growth in the enterprise value of electric car maker Tesla and looked at how that is related to attention in our longitudinal social media data. 

\subsection*{S3 Appendix: detailed analysis of the link between social media attention and enterprise value} \label{S3_Appendix}
In this section we provide the technical details on testing whether there is a granger-causal relation between online social media attention (Reddit and Twitter) and the enterprise value.
Our analysis consists of two steps: first, testing the stationarity of the series and, second, performing the granger-causality test and determining the optimal lag value.
 
\textbf{Step 1: Stationarity and co-integration tests} 
To test the robustness of statistical forecasting using these data, we perform stationarity and co-integration on each of the data series shown in \cref{fig:fig7} (Reddit, Twitter, and Enterprise Value). 
We leverage an augmented Dickey-Fuller test (ADF) to test stationarity, in which the null hypothesis is that it exists a unit root in a time series (i.e., no stationarity) and the alternate hypothesis that the time series is stationary. 
\nameref{S2_Table} summarises the ADF statistic values and p-values of the three series related to Tesla.
The results indicate that we cannot reject the null hypothesis for the Reddit and EV series (p-values $> 5\%$), and only the Twitter series is stationary. 
We transform the series by computing the difference between adjacent values.
Consequently, each series is transformed into the series of `difference from the previous month'.
Applying the same ADF testing shows that the differences series are all stationary (see \nameref{S2_Table}).
However, there are no long-term, predictable patterns for a stationary time series, and the co-integration test could help analyse the original non-stationary time series. 
Engle-Granger two-step method is called to test the stationarity of error term from the linear combination of those variables. 
The social media attention series (Reddit and Twitter) and the enterprise value series are not co-integrated based on the results in \nameref{S2_Table}. 
Based on the above two tests on Tesla, the stationary difference time series are used for the following analysis.

\textbf{Step 2: Granger Causality and VAR Model.} 
Granger causality test is used to examine whether social media data (Reddit \& Twitter) can be used to forecast the changes in enterprise value. 
The null hypothesis is that accounting for lagged values of the social media series does not add explanatory power in predicting the enterprise value of Tesla. 
According to the results in \nameref{S3_Table}, we can reject the null hypothesis and conclude that the Reddit series is granger-causal for the enterprise value of Tesla for lags between 2 and 12 months.
Similarly, the Twitter series has granger causality relation with Tesla EV for lags between 4 and 12 months. 
Based on the results, we can conclude that the fluctuations of social media datasets (such as link counts of Reddit and Twitter) explains some changes of enterprise value of Tesla during the period (2016 Jan and 2019 Sep).

%Considering that Twitter is a more representative social media platform (given its wide spread usage), 
We explore the `optimal' lag in the Reddit and Twitter series to best predict the EV series. 
The Vector Auto-Regression (VAR) model forecasts multiple time series variables using a regression model through their lagged vectors.
We use VAR to determine the optimal lag for predicting the time series of Tesla by selecting the one with the lowest AIC criterion out of the valid options (for which the granger-causality test is significant).
\nameref{S4_Table} summarises the AIC for up to 12 lags, and shows that a lag of 2 months for Reddit and 4 months for Twitter obtain the minimum AIC.
It indicates that the online social media attention is predictive for the enterprise value for Tesla up to four months in advance (for Twitter) and up to one year in advance (for Reddit).

\subsection*{S4 Appendix: Goodness of fit for long-tail distributions} \label{S4_Appendix}
The total online attention of social media from different platforms (Reddit and Twitter) demonstrate different trends (see \nameref{S2_Fig}). The online volume of Reddit steadily increases after 2011 while there is a peak of attention at Twitter platform around 2016. It also explains the changes of the number of links for domains over time at \cref{fig:fig2}. 

We compare the goodness of fit of three distributions, power law, lognormal and exponential distribution, to find the best description of observed Reddit and Twitter data, respectively. A Python package named "powerlaw" is introduced to fit those distributions based on the estimated minimal distance between the theoretical distribution and the empirical data using the Kolmogorov-Smirnov test (see \nameref{S3_Fig}). The CCDF plots of power-law and lognormal distribution are close to the empirical data, while the plot of the exponential distribution is far away from others. 

Then it also provides the method to measure the log-likelihood ratio between two different referenced distributions, where is a ratio of the probability that the observed data comes from the referenced distribution. Since we apply the logarithmic scale on the ratio, positive ratios mean the probability of first theoretical distribution fit to data is higher than the probability of second distribution. Otherwise, the second distribution is preferred. \nameref{S1_Table} illustrates that most of the log likelihood ratios between power law versus lognormal or exponential distribution over all possible periods for Reddit and Twitter are positive. Even though the ratio of power law versus lognormal distrbution is negative in 2016 and 2017, the overall conclusion that the power law distribution is preferred to describe the observed social media data is consistent overall.

\newpage
\paragraph*{S1 Fig.} \label{S1_Fig}
{\bf Popularity in Twitter does not always correspond to the broader population online.} 
In this plot, we show the top domains by linked frequency within Twitter by month. 
Those shaded in dark green are Arabic-language sites that perhaps are automated prayer reminder services. The light grey sites are automated re-tweeting services to help repost material from other social platforms such as Facebook. 
While popular in the number of links appearing on Twitter - these sites are not popular on the broader web. 
The right-hand column with blue bars indicates the Quantcast Global ranking 
%(a respected industry ranking of relative number of visitors to websites worldwide) 
in of each of the domains in the last column for January 2017. Quantcast is an online audience measurement service that covers over 100 million web destinations and estimates relative visitor traffic of all major global websites. The longer the blue bar, the higher the Quantcast rank, i.e. the lower the global popularity.  Many of the webs most popular sites are present: Facebook (ranked 2nd); Youtube (4th) and Instagram (55th) however many of the high ranking sites by the number of links are not also popular on the broader web.
\\[24pt]
%% MAR: comment out the figures
\includegraphics[width=\linewidth]{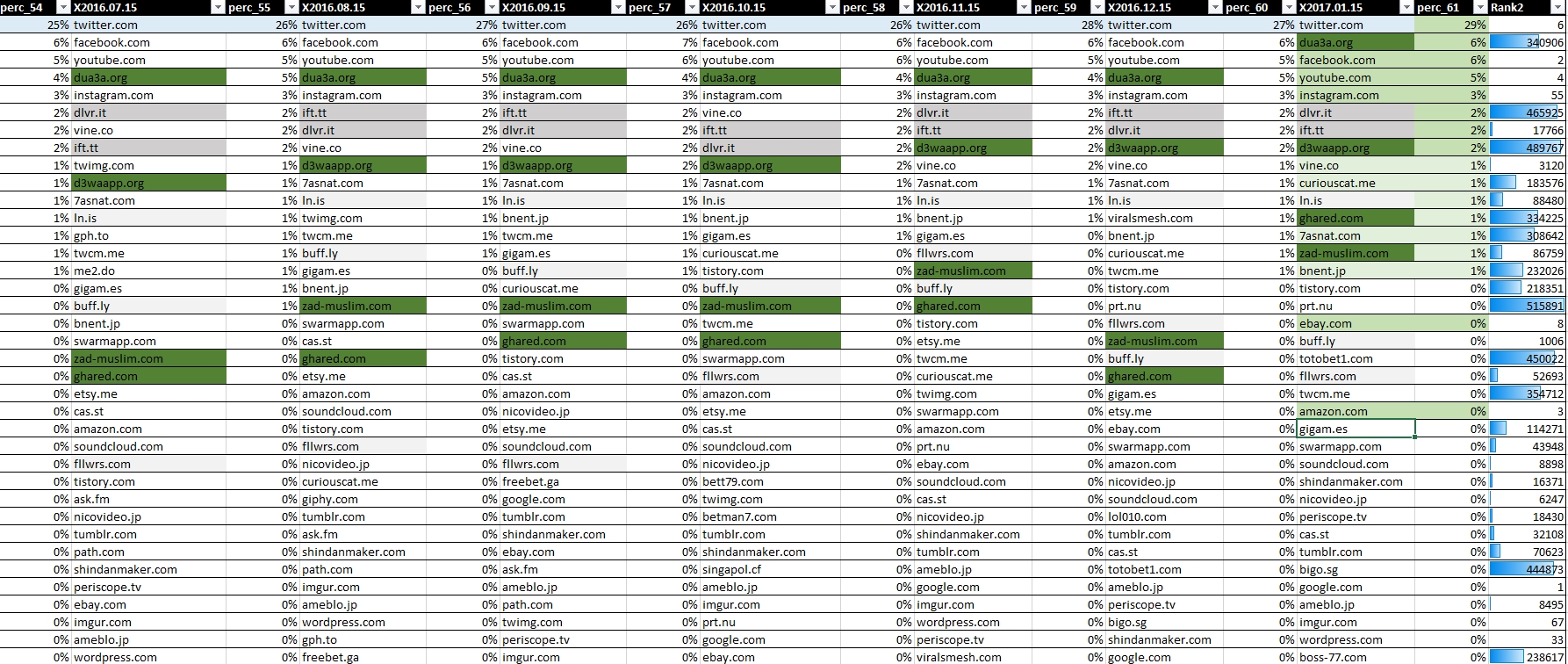}

%\newpage
\paragraph*{S2 Fig.} \label{S2_Fig}
\textbf{Total online attention of Reddit and Twitter vary over time.} 
\textbf{(A)} The monthly volume of links Reddit constantly increases from 2006 to 2019. 
\textbf{(B)} The monthly volume of links in Twitter (measured from 2011 to 2019) peaks in 2017 and decays ever since. The large variations are due to data crawling errors (i.e. periods in which our crawler does not record any data).
The shaded area shows the 95\% confidence level interval for predictions from a linear model.
\\[24pt]
%% MAR: comment out the figures
\includegraphics[width=0.9\linewidth]{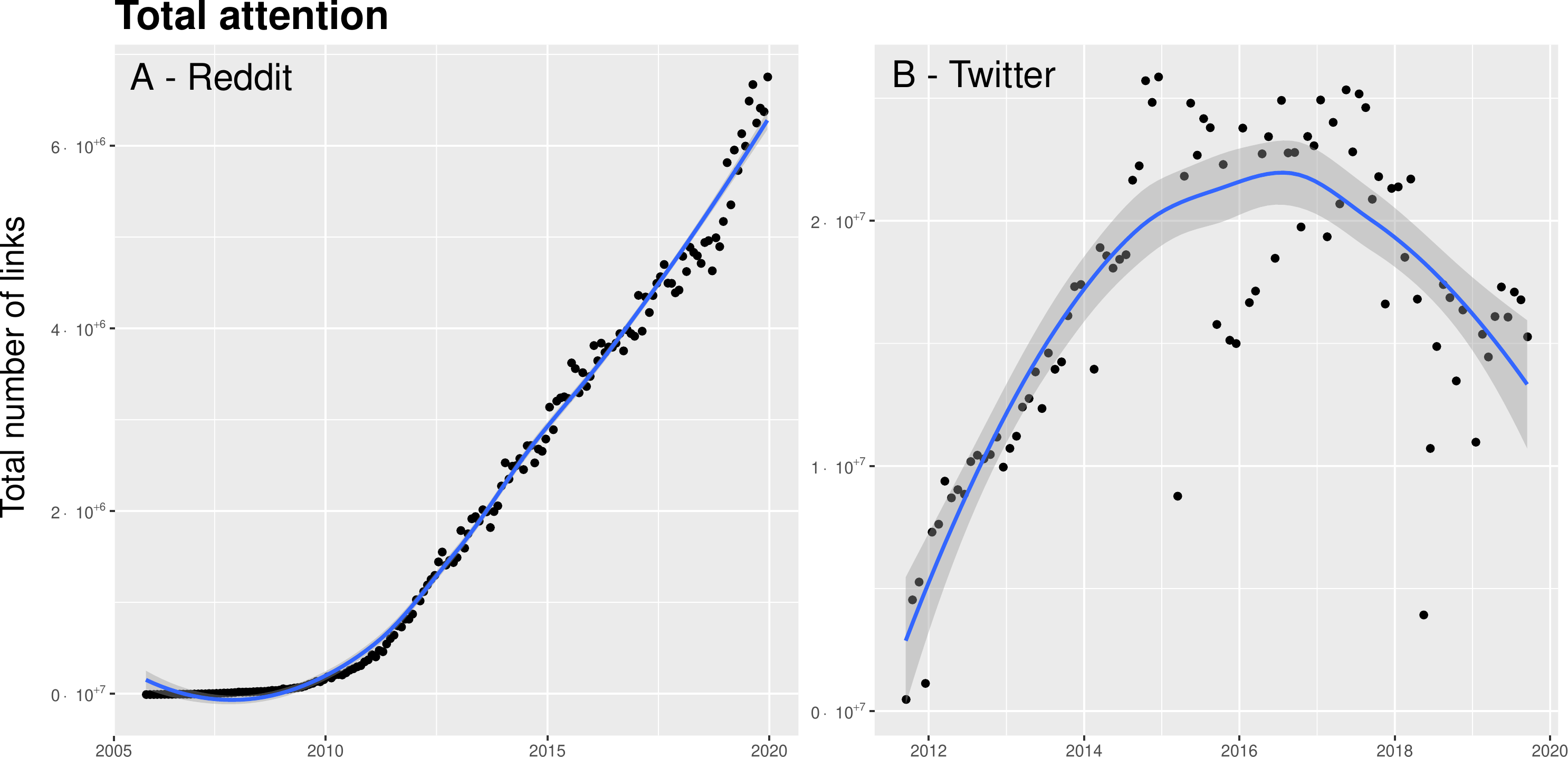}

\newpage
\paragraph*{S3 Fig.}\label{S3_Fig}
\textbf{Comparison of three distributions (i.e. power-law, lognormal and exponential distribution) fitted to Social Media (Reddit and Twitter) in 2016.} 
The log-log plots of the empirical and fitted Complementary Cumulative Distribution Function (CCDF) of the number of links associated with domains in Reddit \textbf{(A)} and in Twitter \textbf{(B)}. 
Black solid lines represent the CCDF of observed data while other dashed lines represent the power-law (red), lognormal (green) and exponential (blue) fitted distribution.
\\[24pt]
%% MAR: comment out the figures before submission
\includegraphics[width=0.99\linewidth]{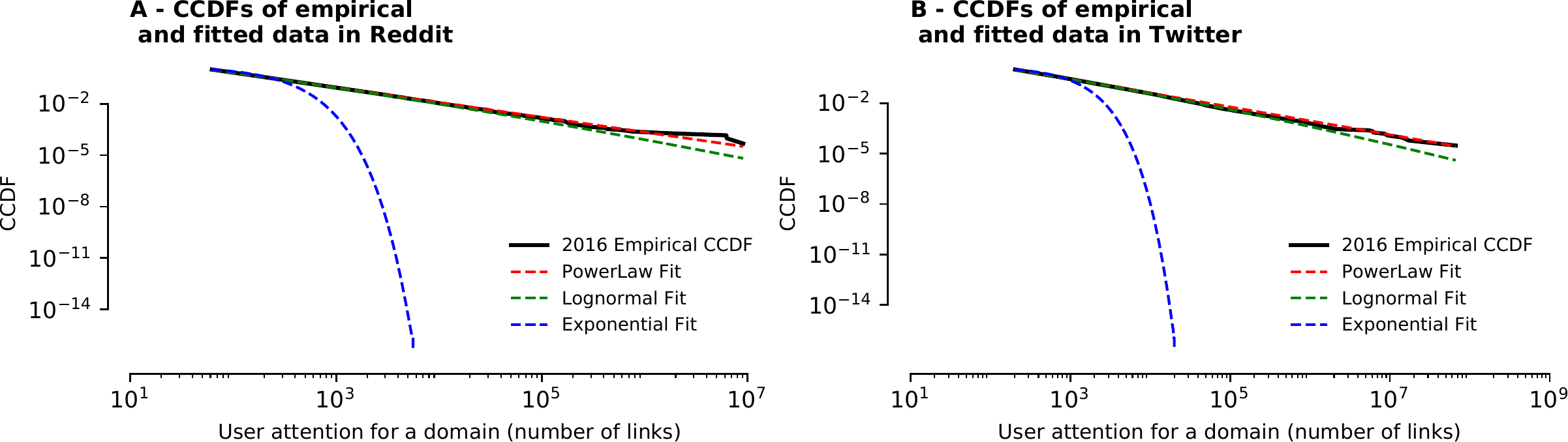}

%\newpage
\paragraph*{S4 Fig.}\label{S4_Fig}
\textbf{The overwhelming majority of the top most visited web sites in the world (according to the Alexa ranking) are included in our data collection.} 
\textit{(left)} 95 of the top 100 are present in our Reddit collection; those missing are concatenations or short-form websites such as Bit.ly, as well as a small number of Advertising network sites and Adult sites that are popular but not shared. 
\textit{(right)} The top Alexa websites account for a greater proportion of all Reddit links over time. 
The Top 1000 ranked websites accounted for less than 60\% of all posts in 2010 and almost 70\% in 2016.
\\[24pt]
%% MAR: comment out the figures
\includegraphics[width=0.95\linewidth]{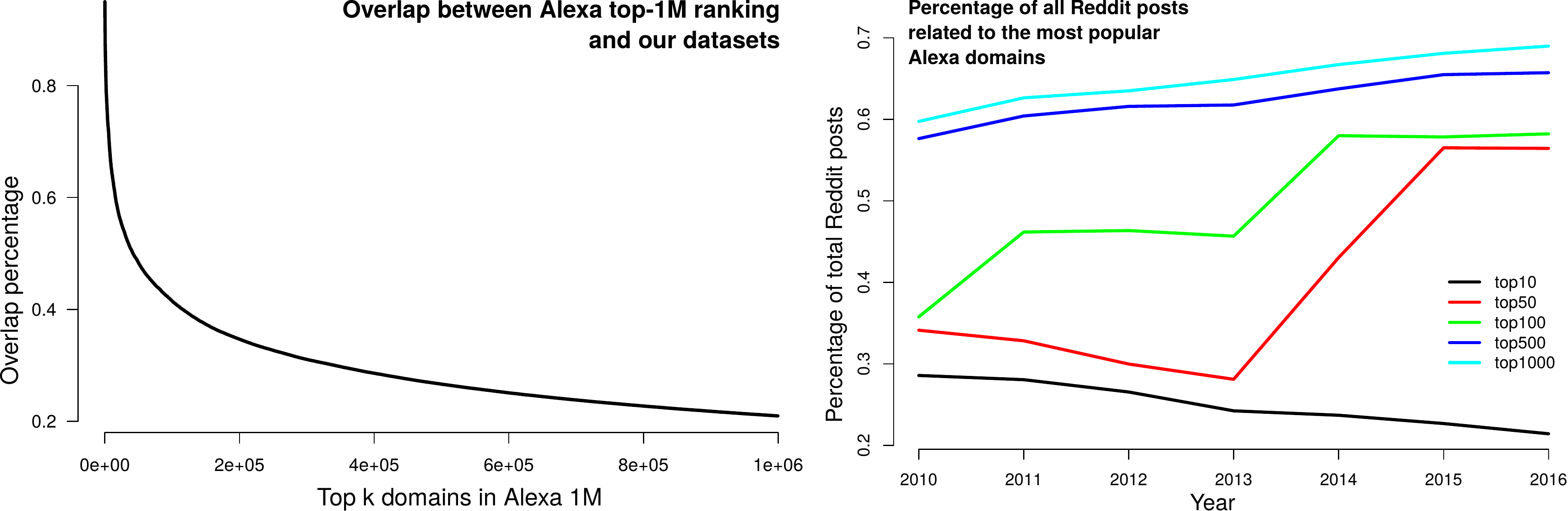}

\newpage
\paragraph*{S5 Fig.}\label{S5_Fig}
\textbf{Online attention for major car manufacturers. Tesla overtakes its competition.}
\textbf{(A)} Wikipedia Page View per month. We collect the monthly page views of the carmakers' pages, from 2017 until 2021.
The y-axis shows the page views, and visibly Tesla's page is viewed consistently more than its competitors.
\textbf{(B)} Google searches in the US.
We use Google search trends to plot Google searches in the US for car manufacturers from 2004 until 2021.
The y-axis shows the search volume in each month, in percentage, relative to the highest value recorded for any of the manufacturers (here, Toyota in Sept 2019).
Visibly, Tesla has consistently increased since 2013 and had overtaken all competitors except Toyota.
\\[24pt]
%% MAR: comment out the figures
\includegraphics[width=0.99\linewidth]{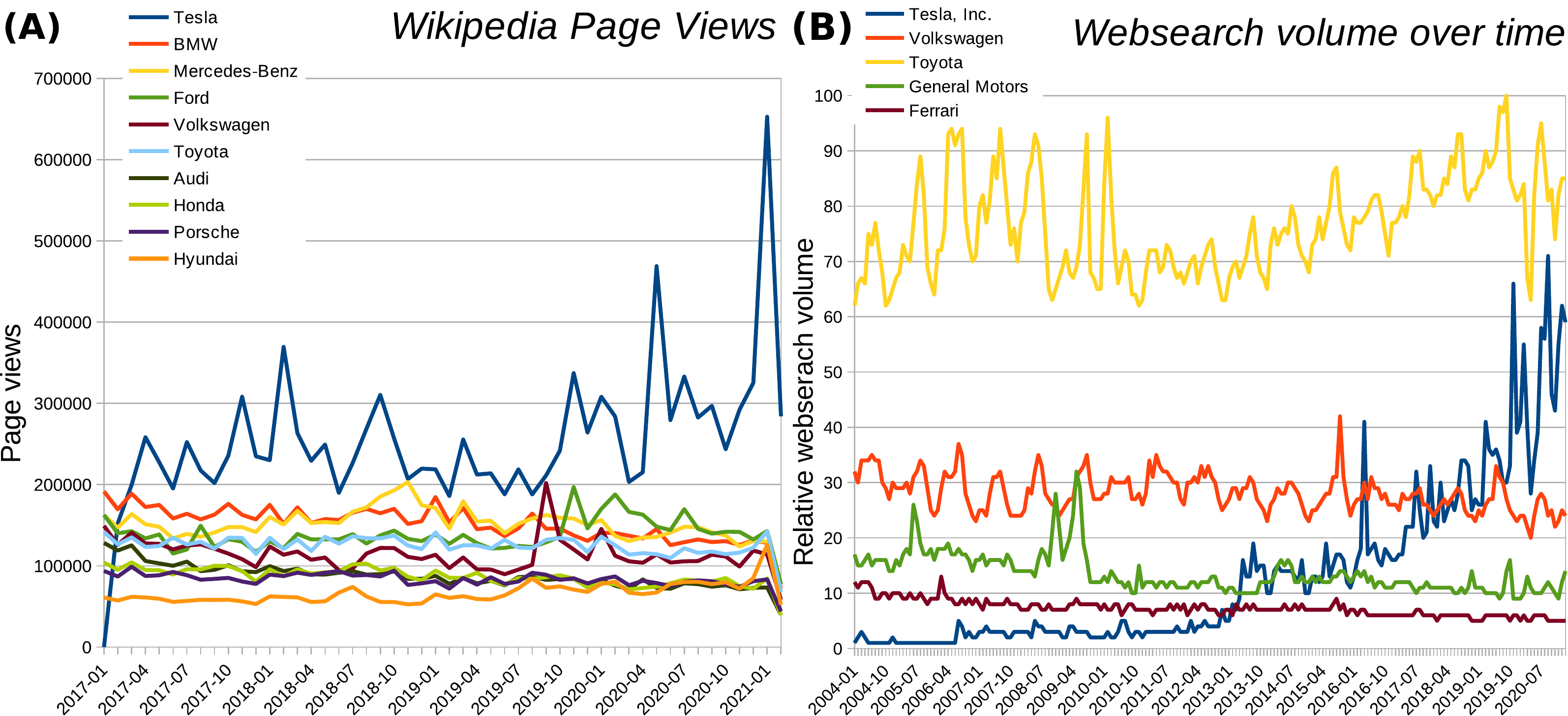}

% \newpage
\paragraph*{S1 Table} \label{S1_Table}  
\textbf{Log likelihood ratios of power-law vs. lognormal and exponential distribution, for Reddit and Twitter.} 
Positive ratios indicated that power-law is prefered to describe Social Media data.
Conversely, negative values indicate that lognormal/exponential distribution is preferred.
{\centering
    \begin{tabular}{lrrrrr}
        \toprule
         Year & \multicolumn{2}{c}{Reddit} &   & \multicolumn{2}{c}{Twitter}    \\ 
        \cline{2-3} \cline{5-6} 
      &   \makecell{Power law \\ vs. \\  Lognormal} & \makecell{Power law \\ vs. \\ Exponential} & & \makecell{Power law \\ vs. \\ Lognormal} & \makecell{Power law \\ vs. \\ Exponential} \\     
    %   &   \centered {Power law} & \centered {Power law} & & \centered {Power law} & \centered {Power law} \\ 
    %   &   \centered {VS} & \centered {VS} & & \centered {VS} & \centered {VS} \\ 
    %   &   \centered {Lognormal} & \centered {Exponential} & & \centered {Lognormal} & \centered {Exponential} \\  

        \midrule
         2006   &  1.422   & 1.799 &
            & -   & -         \\       
         2007   &  2.767   & 2.457  & 
            & -   & -          \\         
         2008   &  1.467   & 2.449       & 
            & -    & -          \\
         2009   &  2.933   & 2.690       & 
            & -    & -          \\
         2010  &  1.373   & 2.632      & 
           & -    & -           \\    
         2011   & 2.007    & 2.734       &  
            &  3.637  & 4.288  \\
         2012   & 3.261    & 3.067       & 
             &  1.127  & 3.446   \\
         2013   & 2.655    & 3.026       &   
            &  2.395  & 3.918  \\
         2014   & 1.673    & 2.781       &    
            &  0.203   & 5.767   \\
         2015  & 0.835    & 2.744      &     
           &  0.395   & 3.862  \\
         2016   &  1.552  & 2.874      &  
            &  -2.554  & 2.930   \\
         2017   &  1.765   & 2.771       &  
            &  -6.382  & 2.382   \\
         2018   &  3.892   & 2.789       &  
            &  1.048  & 1.641 \\
         2019   &  2.643   & 2.341       &  
            &  1.618  & 1.476
                    \\\bottomrule
    \end{tabular}
}

\newpage
\paragraph*{S2 Table} \label{S2_Table} 
\textbf{Summary of the stationarity and the co-integration tests, for linking social media attention and enterprise value for Tesla:}
ADF statistics and p-values results show that the Reddit and EV series are not stationary.
However, the first difference of the series (i.e the difference between two adjacent values) is stationary for all variables (Reddit, Twitter and EV).
The co-integration test indicates no co-integration between Reddit and EV, and Twitter and EV respectively.
{\centering
	\begin{tabular}{l|rrrr}
		\toprule
		Tesla & \multicolumn{1}{l}{Time Series} & \multicolumn{1}{l}{ADF Statistics} & \multicolumn{1}{l}{P-value} \\  \hline
        \multirow{6}{*}{Stationarity Test}&\multicolumn{1}{l}{Link counts in Reddit}&-0.19&46.80\% \\ 
        & \multicolumn{1}{l}{First difference Reddit} & -3.20 & 2.01\%* \\ 
        & \multicolumn{1}{l}{Link counts in Twitter} & -6.40 & 0.00\%* \\ 
        & \multicolumn{1}{l}{First difference Reddit} & -4.68 & 0.00\%* \\ 
        & \multicolumn{1}{l}{Enterprise Value} & -1.99 & 28.68\% \\ 
        & \multicolumn{1}{l}{First difference in EV} & -7.48 & 0.00\%* \\ \hline		
        \multirow{2}{*}{Co-integration Test} & \multicolumn{1}{l}{Error term of Reddit \& EV} & -2.09 & 24.91\% \\ 
        & \multicolumn{1}{l}{Error term of Twitter \& EV} & -2.20 & 20.49\% \\ 
		
		\bottomrule
	\end{tabular}
}

% \newpage
\paragraph*{S3 Table} \label{S3_Table} 
\textbf{Summary of the results of Granger-causality test between social media attention and the enterprise value.}
The lag column  indicate the maximum lag of time included in the Granger-causality test.
The Reddit and Twitter columns show the corresponding p-value of Granger test of the respective social media platform and the enterprise value series. 
Values prefixed by asterisk (\textbf{*}) indicate statistical significance with p-value $< 5\%$.
Results indicate that the enterprise value is granger-caused by both Reddit (after a lag of 2) and Twitter (after a lag of 4).
\\
{\centering
	\begin{tabular}{lrrr}
		\toprule
		Lag & \multicolumn{1}{l}{Reddit} & \multicolumn{1}{l}{Twitter} \\ \midrule
		1 month & 24.64\% & 46.07\% \\ 
		2 months & 1.32\%\textbf{*} & 63.71\% \\ 
		3 months & 2.41\%\textbf{*} & 54.62\% \\ 
		4 months & 0.96\%\textbf{*} & 0.30\%\textbf{*} \\ 
		5 months & 0.40\%\textbf{*} & 0.52\%\textbf{*} \\ 
		6 months & 0.03\%\textbf{*} & 0.02\%\textbf{*} \\ 
		7 months & 0.04\%\textbf{*} & 0.04\%\textbf{*} \\ 
		8 months & 0.06\%\textbf{*} & 0.09\%\textbf{*} \\ 
		9 months & 0.00\%\textbf{*} & 0.11\%\textbf{*} \\ 
		10 months & 0.00\%\textbf{*} & 0.08\%\textbf{*} \\ 
		11 months & 0.01\%\textbf{*} & 0.00\%\textbf{*} \\ 
		12 months & 0.00\%\textbf{*} & 0.00\%\textbf{*} \\  		\bottomrule
	\end{tabular}
}

\newpage
\paragraph*{S4 Table} \label{S4_Table} 
\textbf{The value of the AIC criterion of VAR model} between Twitter and Enterprise Value of Tesla, for lags values between 0 and 12 months. 
We chose as the `best lag' the lag that minimises the AIC (indicated by \textbf{*}) from the set of valid values (for which the granger-causality test is significant, shown in \textbf{bold face}).
\\
{\centering
	\begin{tabular}{lrr}
		\toprule
		Lag & \multicolumn{1}{l}{Reddit} & \multicolumn{1}{l}{Twitter} \\ \midrule
		0 & \textit{13.15} & \textit{12.97} \\ 
		1 month & \textit{13.21} & \textit{12.90} \\ 
		2 months & \textbf{13.25} & \textit{12.97} \\ 
		3 months & \textbf{13.36} & \textit{12.84} \\ 
		4 months & \textbf{13.31} & \textbf{12.70*} \\ 
		5 months & \textbf{13.31} & \textbf{12.91} \\ 
		6 months & \textbf{13.29} & \textbf{12.92} \\ 
		7 months & \textbf{13.48} & \textbf{12.94} \\ 
		8 months & \textbf{13.69} & \textbf{13.14} \\ 
		9 months & \textbf{13.83} & \textbf{13.40} \\ 
		10 months & \textbf{13.75} & \textbf{13.49} \\ 
		11 months & \textbf{13.74} & \textbf{13.41} \\ 
		12 months & \textbf{13.01*} & \textbf{13.40} \\  		\bottomrule
	\end{tabular}
}

\newpage
\singlespacing

\paragraph*{S5 Table} \label{S5_Table} 
\textbf{Rivalfox data.} 
The most linked global online service (shown in \textbf{bold face}) in twelve of the functional categories defined by Crunchbase, and their three main direct rivals identified by Rivalfox.
\small{
\begin{tabular}{lllr}
\toprule
Category & Domain & Company & \multicolumn{1}{l}{Year started} \\ \midrule
Video & \url{vimeo.com} & Vimeo & 2004 \\ 
Video & \url{www.truveo.com} & Truveo & 2004 \\ 
Video & \url{www.dailymotion.com} & Dailymotion & 2005 \\
Video & \url{www.youtube.com} & \textbf{Youtube} & 2005 \\ 
Filesharing & \url{drive.google.com} & Google Drive & 2012 \\ 
Filesharing & \url{home.elephantdrive.com} & Elephant Drive & 2005 \\ Filesharing & \url{www.sugarsync.com} & Sugarsync & 2004 \\ 
Filesharing & \url{www.dropbox.com} & \textbf{Dropbox} & 2007 \\ 
Accommodation & \url{www.homeaway.com} & HomeAway & 2005 \\ 
Accommodation & \url{www.wimdu.com} & Wimdu & 2011 \\ 
Accommodation & \url{www.9flats.com} & 9flats & 2012 \\ 
Accommodation & \url{www.airbnb.com} & \textbf{AirBnb} & 2008 \\ 
Music streaming & \url{hello.simfy.de} & simfy & 2006 \\
Music streaming & \url{www.rdio.com} & Rdio & 2008 \\ 
Music streaming & \url{www.deezer.com} & Deezer & 2006 \\ 
Music streaming & \url{www.spotify.com} & \textbf{Spotify} & 2008 \\ 
Ride sharing & \url{www.lyft.com} & Lyft & 2007 \\ 
Ride sharing & \url{www.hailoapp.com} & Hailo & 2010 \\
Ride sharing & \url{www.side.cr} & Sidecar & 2010 \\ 
Ride sharing & \url{www.uber.com} & \textbf{Uber} & 2009 \\
Search & \url{www.bing.com} & Bing & 2009 \\ 
Search & \url{www.duckduckgo.com} & DuckDuckGo & 2008 \\ 
Search & \url{www.yahoo.com} & Yahoo! & 1994 \\ 
Search & \url{www.google.com} & \textbf{Google} & 1998 \\ 
Social Network & \url{twitter.com} & Twitter & 2006 \\
Social Network & \url{plus.google.com} & Google Plus & 1998 \\
Social Network & \url{tumblr.com} & Tumblr & 2007 \\
Social Network & \url{facebook.com} & \textbf{Facebook} & 2004 \\
General Retail & \url{amazon.com} & \textbf{Amazon} & 1994 \\
General Retail & \url{walmart.com} & Walmart & 1962 \\
General Retail & \url{ebay.com} & eBay & 1995 \\
General Retail & \url{bestbuy.com} & Best Buy & 1966 \\
Movies & \url{Veed.me} & Veed & 2013 \\
Movies & \url{hulu.com} & Hulu & 2007 \\
Movies & \url{netflix.com} & \textbf{Netflix} & 1997 \\
Movies & \url{amazon.com/video} & Amazon video & 2006 \\
Ephemeral messaging & \url{wickr.com} & Wickr & 2011 \\
Ephemeral messaging & \url{clipchat.com} & Clipchat & 2013 \\
Ephemeral messaging & \url{sling.me} & Slingshot & 2014 \\
Ephemeral messaging & \url{snapchat.com} & \textbf{Snapchat} & 2011 \\
Action cameras & \url{contour.com} & Contour & 2004 \\
Action cameras & \url{eyesee360.com} & EyeSee360 & 1998 \\
Action cameras & \url{vievu.com} & Vievu & 2007 \\
Action cameras & \url{gopro.com} &  \textbf{GoPro} & 2003 \\
Dating app & \url{chatimity.com} &  Chatimity & 2011 \\
Dating Apps & \url{hinge.co} &  Hinge & 2011 \\
Dating Apps & \url{wyldfireapp.com} &  Wyldfire & 2013 \\
Dating Apps & \url{gotinder.com} &  \textbf{Tinder} & 2012 \\
\bottomrule
\end{tabular}
}

\end{document}